\documentclass[final,english,twocolumn,amssymb,aps,prb,
longbibliography]{revtex4-2}
\usepackage{graphicx}
\usepackage{natbib}
\usepackage{amsmath}
\usepackage{amssymb}
\usepackage{appendix}
\usepackage{bm}
\usepackage{overpic}
\usepackage[dvipsnames]{xcolor}{\huge }
\usepackage[section]{placeins}
\definecolor{darkblue}{rgb}{0,0,.65}
\definecolor{darkgreen}{rgb}{0.28,0.41,0.19}

\usepackage{multirow}

\usepackage{nicefrac}
\usepackage[%
    pdfauthor={Imre Hagymasi},%
  pdfstartview=FitH,%
  breaklinks=true,%
  bookmarks=true,%
  colorlinks=true,%
  anchorcolor=black,%
  citecolor=blue,
  filecolor=black,%
  menucolor=black,%
  urlcolor=darkblue,%
  linkcolor=blue,%
 ]{hyperref}
\usepackage[all]{hypcap} %

\usepackage[capitalize]{cleveref}

\newcommand{\cf}{\textit{cf.} }

\newcommand{\ih}[1]{\textcolor{Black}{#1}}

\graphicspath{{images/}}
\begin{document}

\title{Magnetic phases of the periodic Anderson model in two dimensions}

\author{Imre Hagym\'asi}
\affiliation{Institute 
for Solid State
Physics and Optics, Wigner Research Centre for Physics, Budapest H-1525 P.O. 
Box 49, Hungary
}
\date{\today}

\begin{abstract}
We investigate the ground-state properties of the periodic Anderson model on the square lattice across various band fillings. Employing the infinite projected entangled-pair states (iPEPS) technique, we can determine the magnetic ground states accurately and compare them to mean-field predictions to highlight the effects of quantum fluctuations. At half-filling, we analyze the transition between the antiferromagnetic and paramagnetic (Kondo singlet) phases as a function of hybridization and $f$-level energy, finding excellent agreement with existing quantum Monte Carlo studies in the case of hybridization. For $n = 1.5$ electrons per site, we identify a novel correlated antiferromagnetic diagonal stripe phase as the ground state, which competes with its ferromagnetic partner state.
\end{abstract}

\maketitle

\section{Introduction}
In $f$-electron systems, we frequently encounter intriguing phenomena such as unconventional superconductivity, heavy-fermion behavior, and mixed-valence states \cite{Hewson_1993,fazekas_book_1999}. The heavy-fermion behavior is a consequence of the greatly enhanced density of states due to strong electron correlations but the state usually remains paramagnetic. This is somewhat counterintuitive since according to the Stoner criterion electrons with a high density of states at the Fermi energy should be highly prone to magnetic ordering. Indeed, many heavy-fermion compounds exhibit antiferromagnetic order at very low temperatures, for example U$_2$Zn$_{17}$, UCd$_{11}$ \cite{stewart_review}. However, there are exceptions, such as UCu$_5$, where magnetic ordering coexists with heavy-fermion behavior \cite{ott_review}. While the above examples pertain to three-dimensional systems, recent advancements in synthesis techniques have enabled the creation of two-dimensional $f$-electron systems. Notably, heterostructures involving Eu or Gd provide a platform to explore the interplay of magnetism, strong correlations, and enhanced quantum effects in reduced dimensions \cite{2d_review}. These systems may open up new avenues for studying heavy-fermion behavior in lower-dimensional geometries, where quantum fluctuations can play an even more significant role.
\par The periodic Anderson model (PAM) is widely regarded as the minimal model that captures the essential physics of these materials. It describes the interaction between localized, strongly correlated electrons and delocalized conduction electrons from a broad energy band. The Hamiltonian is expressed as:
\begin{equation}   \begin{split}
     \mathcal{H} = & \sum_{\bm{k},\sigma}\varepsilon_c(\bm{k})
       \hat{c}_{\bm{k}\sigma}^{\dagger}
           \hat{c}^{\phantom \dagger}_{\boldsymbol{k}\sigma}
           +\varepsilon_f\sum_{j,\sigma}\hat{n}^f_{j\sigma} \\
           & - V\sum_{j,\sigma}(\hat{f}_{j\sigma}^{\dagger}
          \hat{c}^{\phantom \dagger}_{j\sigma}
    +\hat{c}_{j\sigma}^{\dagger} \hat{f}^{\phantom \dagger}_{j\sigma})
     +U_f\sum_{j}\hat{n}^f_{j\uparrow} \hat{n}^f_{j\downarrow} .
\label{PAM:Hamiltonian}
\end{split}
\end{equation}
Here, $\hat{c}_{\bm{k}\sigma}^\dagger$ ($\hat{c}_{\bm{k}\sigma}$) represents the creation (annihilation) operator for conduction electrons with wave vector $\bm{k}$ and spin $\sigma$. Similarly, $\hat{f}_{j\sigma}^\dagger$ ($\hat{f}_{j\sigma}$) denotes the creation (annihilation) operator for localized $f$-electrons at site $\bm{r}_j$. The particle number operators for $f$- and $c$-electrons are given by $\hat{n}^f_{j\sigma} = \hat{f}_{j\sigma}^\dagger \hat{f}_{j\sigma}$ and $\hat{n}^c_{j\sigma}=\hat{c}_{j\sigma}^\dagger \hat{c}_{j\sigma}$, respectively. The hybridization amplitude between the $f$- and $c$-electron states is denoted by $V$, while $U_f$ represents the on-site Hubbard repulsion between $f$-electrons.
 The average number of $c$- and $f$-electrons per site, denoted as $n_c$ and $n_f$, respectively, can range from zero to two. The model's filling is defined as the ratio of the total electron density per site ($n = n_c + n_f$) to the maximum possible electron occupancy ($n_{\text{max}} = 4$). For a comprehensive overview of the fundamental properties of this model, refer to Ref.~\cite{noce_review}.
\par In spite of its apparent simplicity, the PAM has no analytic solution even in one dimension unlike the Hubbard model, though exact solutions can be constructed in certain cases \cite{gulacsi_prl,gurin_2001prb}. Therefore other techniques have been used to study this model, which often involve approximations. In one dimension the density-matrix renormalization group provides a quasi error-free description of the PAM \cite{noack_prb2001,santos_2011prb, hagymasi_2014prb}. At half filling a gapped ground-state emerges with antiferromagnetic correlations between the $f$-electrons mediated by the RKKY interaction. In two dimensions the magnetic properties of the PAM has been extensively explored using quantum Monte Carlo \cite{vekic_prl}, variational Monte Carlo methods \cite{watanabe_2009, kubo_2015} as well as dynamical vertex approximation \cite{schafer_2019}. In three dimensions, the model has been investigated with mean-field theory \cite{spalek_prb} and, on the Bethe and cubic lattices, with dynamical mean-field theory \cite{nolting_prb, koga_prb,kotliar_2008prl,vondelft_2024prx}. While dynamical mean-field theory (DMFT) yields exact results in infinite spatial dimensions, it remains uncertain how quantum fluctuations in two  dimensions modify these findings.  The quantum Monte Carlo technique provides reliable results for the half-filled and symmetric ($\varepsilon_f=-U_f/2$) case, however, moving away from the symmetric case or considering other fillings introduce the sign problem \cite{troyer_prl}, which can significantly complicate or even hinder the simulations. To mitigate the sign problem the constrained path quantum Monte Carlo technique has been successfully used to study the PAM at various fillings \cite{bonca_1998prb,batista_2001prb} but it faces scaling challenges with the system size. Variational techniques have also been widely applied to study the properties of the PAM. These methods either use the Monte Carlo technique \cite{watanabe_2009,kubo_2015} or the Gutzwiller approximation \cite{fazekas_1987,gebhard_prb1991,itai_prb,hagymasi_2012prb} to evaluate expectation values. While the former case provides an exact evaluation of the observables and the calculation is only impacted by the variational Ansatz, the latter case involves a further mean-field like approximation as well. Later on, the PAM has been investigated using the density-matrix embedding theory near half filling \cite{Yang_2018}. Consequently, the analysis of the PAM with techniques that do no rely on rigid Ans\"atze or uncontrolled approximations is highly warranted.
\par The goal of this paper is to investigate the ground-state properties of the periodic Anderson model (PAM) on a square lattice at various fillings as a function of the hybridization strength and the \(f\)-level energy. We employ the infinite projected entangled-pair states (iPEPS) technique \cite{cirac_revmod,verstraete2004,nishio2004,cirac_2008prl,corboz_2010prb,nishino_2001}, which overcomes the aforementioned challenges and has been successfully applied to study frustrated spin systems \cite{peschke_2022prb,corboz_2014prl,liao_2017prl,schmoll_2023prb} as well as correlated fermionic systems \cite{zheng_2017science,corboz_tJ_2014prl,Lee2020} in two dimensions. Furthermore, we compare our results to those obtained via mean-field theory to emphasize the role of quantum fluctuations. In the symmetric half-filled case, we observe that increasing the hybridization suppresses the antiferromagnetic order, and beyond a critical value, a Kondo singlet phase emerges. Our estimate for the critical point aligns excellently with previous quantum Monte Carlo studies. The antiferromagnetic order persists when the \(f\)-level energy is increased; however, it is suppressed around $n_f \sim 0.5$ occupancy, where a paramagnetic state appears -- significantly deviating from the predictions of mean-field theory.
Our most striking observation is the emergence of a diagonal stripe state for $n=1.5$ filling hosting a correlated antiferromagnetic order.  This state strongly competes with its ferromagnetic counterpart in certain parameter regimes.  In general, our studies demonstrate that the iPEPS technique can produce reliable results for challenging parameter regimes and fillings of a correlated two-band model.
\par The paper is organized as follows. In Sec.~II.~we review the methodological details. In Sec.~III.~A and B we present our results for the half-filled and $n=1.5$ filling case, respectively. Finally, in Sec.~IV.~ we conclude our findings.

\section{Methods}
We use the iPEPS method \cite{cirac_revmod,verstraete2004,nishio2004,cirac_2008prl,corboz_2010prb, nishino_2001} to study the ground-state properties of the PAM on the square lattice, thus $\varepsilon_c({\bm{k}})=-2t(\cos( k_x) + \cos( k_y))$ and we set the hopping amplitude $t$ to $t=1$ and the lattice constant to unity. The iPEPS technique is a variational method that represents the wave function as a network of rank-5 tensors. For a detailed review of the method, see Refs.~\cite{cirac_revmod,verstraete2004,nishio2004,cirac_2008prl,corboz_2010prb,nishino_2001,bruognolo_scipost2021} \ih{and the Appendix}. The ground state is assumed to be describable using a supercell of tensors that tiles the entire lattice. Each tensor has four auxiliary legs, which connect to neighboring sites, and one physical leg. The accuracy of the approach is controlled by the bond dimension, $D$, of the auxiliary legs. By gradually increasing $D$, the Ansatz can incorporate more entanglement, improving the representation of the ground state. Each tensor contains $dD^4$ variational parameters, where $d$ is the dimension of the local Hilbert space. In the periodic Anderson model (PAM), two fermionic degrees of freedom per lattice site are present. To adapt to the algorithm, these are merged into a supersite with a local Hilbert space dimension of $d = 16$, allowing the originally developed iPEPS techniques to be directly applied.
We optimize the tensors using the fast full update (FFU) algorithm, employing gauge fixing to enhance convergence \cite{orus_ffu_2015prb}, and compute observables using the corner transfer matrix (CTM) technique \cite{nishino_ctm,orus_ctm}. The boundary bond dimension, $\chi$, which determines the accuracy of tensor contractions in the CTM method, was set to $\chi = D^2$. Larger $\chi$ values were tested, but their effects were negligible compared to the dependence on $D$ and did not introduce significant quantitative changes to the results. In our simulations, we utilized $\mathrm{U(1)}$ spin and $\mathrm{U(1)}$ charge symmetries to reduce the sizes of the dense tensor blocks. In certain cases, only the $\mathrm{U(1)}$ charge symmetry was applied. The use of both symmetries allowed us to reach bond dimensions up to $D = 11$, while simulations using only the $\mathrm{U(1)}$ charge symmetry were limited to $D = 9$. We considered supercells with sizes $2\times 2$, $2\times 4$ and $4\times 4$. Observables were extrapolated to the error-free limit ($D \to \infty$) as a function of $1/D$. For the energies, we employed extrapolation as a function of the normalized cost function with the time step of the imaginary time evolution, $w$, which has been shown to provide higher accuracy \ih{(see Appendix for more details).} This approach yields smoother energy curves as a function of $w$ compared to $1/D$ \cite{corboz_error}. We estimate the extrapolation error as half the distance between the last data point and the extrapolated value.
\par We also perform mean-field calculations using the same supercell sizes as in the iPEPS calculations allowing arbitrary magnetic ordering within the supercell generalizing the approach of Ref.~\cite{hagymasi_2013prb}. We use the standard decoupling of the Hubbard term $\hat{n}^f_{j\uparrow} \hat{n}^f_{j\downarrow} \sim \hat{n}^f_{i\uparrow}\langle \hat{n}^f_{i\downarrow}
\rangle+\hat{n}^f_{i\downarrow}\langle \hat{n}^f_{i\uparrow} \rangle -\langle
\hat{n}^f_{i\downarrow}
\rangle \langle \hat{n}^f_{i\uparrow} \rangle$ and solve the single-particle problem in a self-consistent way.

\section{Results}
\subsection{Half-filled case}
\begin{figure}[t]
    \centering
\includegraphics[width=\columnwidth]{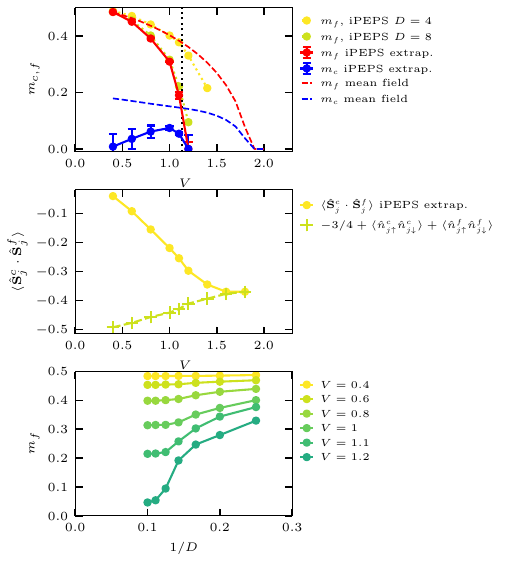}
  \caption{The sublattice magnetizations, $m_{c,f}$, (upper panel) and local spin correlation between conduction and $f$ electrons (middle panel) as a function of the hybridization in the symmetric ($\varepsilon_f=-U_f/2$, $U_f=4$) half-filled model. For the magnetizations we plotted the mean-field results as well. The bottom panel shows the sublattice magnetization of the $f$ electrons as a function of the inverse bond dimension. The dotted vertical line on the top panel denotes the quantum Monte Carlo estimate for the critical point.} 
  \label{fig:sym_case}
\end{figure}

First, we consider the half-filled model at the symmetric point $\varepsilon_f=-U_f/2$, which ensures that the electron density remains for both types of electrons $ n_f =n_c =1$ (\ih{Mott-Hubbard regime}). This case has been investigated in literature with various many-body techniques like quantum Monte Carlo \cite{vekic_prl} or the dynamical vertex approximation \cite{schafer_2019} and we choose $U_f=4$ so that we can compare our results directly to them. \ih{Since the quantum Monte Carlo method is free from the sign problem at this point, this serves also as an independent benchmark for the iPEPS technique.} Based on the Doniach's phase diagram, we expect that for small values of the exchange coupling, $J\sim V^2/U_f$, the RKKY interaction is dominant, which mediates antiferromagnetic fluctuations between the $f$ electrons and leads to magnetic ordering, while for large $J$ the Kondo effect prevails and local singlets are formed. To capture the antiferromagnetic order we use a $2\times 2$ supercell in the iPEPS simulation. The sublattice magnetizations for the conduction ($m_c$) and $f$ electrons ($m_f$) are defined in the following way:
\begin{gather}
    m_{c,f}=\frac{1}{N}\sum_i (-1)^{\alpha_i} m_i^{c,f}, \\
    m_i^{c,f}=\frac{1}{2}\langle \hat{n}^{c,f}_{i\uparrow}-\hat{n}^{c,f}_{i\downarrow}\rangle 
\end{gather}
where $m_i^{c,f}$ denotes the local magnetic moments and the summation extends over the unit cell, $N$ is the number of the sites of the supercell, that is, $N=4$ and $\alpha_i=1$ and $-1$ for the $A$ and $B$ sublattices, respectively. The results for these quantities are shown in the upper panel of Fig.~\ref{fig:sym_case}.
The iPEPS results clearly indicate that the antiferromagnetic phase disappears around $V_{\rm crit}\sim1.2$, which is in excellent agreement with the quantum Monte Carlo result, $V_{\rm crit}\sim1.1(1)$ \cite{vekic_prl}, while the dynamical vertex approximation estimates it as $V_{\rm crit}\sim 0.91$ \cite{schafer_2019}. For small hybridizations, \(V \lesssim 0.5\), the conduction electron magnetization \(m_c\) is indistinguishable from zero, while \(m_f\) is nearly maximal. \ih{For $V=1$ the quantum Monte Carlo study predicts $m_f\sim0.26(5)$, which is again in perfect agreement with the iPEPS value of $m_f\sim0.3$.} The behavior of \(m_c\) resembles that observed in the Kondo lattice model on the square lattice \cite{assaad_klm_prl}. Moreover, it can be clearly seen from Fig.~\ref{fig:sym_case} (bottom panel) that larger and larger bond dimensions are required as we approach $V_{\rm crit}\sim1.2$, \ih{which confirms the increasing correlation length in the system.} Above the critical point, the conduction and \(f\)-electrons form a local singlet. This is evident from the local \(\langle \hat{\bm{S}}_j^c \cdot \hat{\bm{S}}_j^f\rangle \) correlator, where
\begin{equation}
    \hat{\bm{S}}_j^{c,f} = \frac{1}{2} \sum_{\alpha,\beta} \hat{a}_{j\alpha}^\dagger \vec{\sigma}_{\alpha\beta} \hat{a}_{j\beta}, \quad a\in \{c,f\}
\end{equation}
and $\vec{\sigma}_{\alpha\beta}$ represents the vector of Pauli matrices acting on the spin indices $\alpha$ and $\beta$. This correlator approaches  $-3/4+\langle \hat{n}^f_{j\uparrow} \hat{n}^f_{j\downarrow} \rangle + \langle \hat{n}^c_{j\uparrow} \hat{n}^c_{j\downarrow} \rangle$ corresponding to the condition $\langle (\hat{\bm{S}}_j^c+\hat{\bm{S}}_j^f)^2 \rangle=0$. \ih{Since the sublattice magnetization decreases continuously to zero the transition is of second order between the two phases in agreement with previous studies \cite{vekic_prl,assaad_klm_prl}. }
\par It is also worth comparing these results to the mean-field predictions. Although the mean-field theory exhibits qualitatively similar behavior, it significantly overestimates the location of the critical point and the magnetization of conduction electrons, indicating that it fails to capture the itinerant magnetization accurately.  
\par \ih{Having established that iPEPS reproduces the quantum Monte Carlo results} for the symmetric model we now investigate what happens if the $f$-level energy is increased and the system enters the mixed-valence regime. 
\begin{figure}[t]
    \centering
\includegraphics[width=\columnwidth]{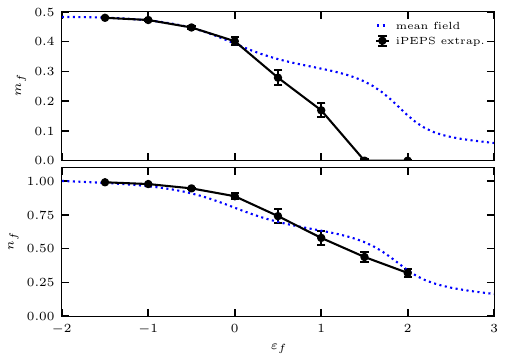}
  \caption{The sublattice magnetization of the $f$ electrons (upper panel) and the $f$ level occupancy (bottom panel) as the function of the $f$-level energy for the parameters $U_f=4$, $V=0.4$ and $n=2$. The symbols are the extrapolated ($D\rightarrow\infty$) iPEPS results and the dotted curve corresponds to the mean-field solution. }
  \label{fig:mixed_valence_case}
\end{figure}
Fig.~\ref{fig:mixed_valence_case} clearly shows that the magnetic moments decrease as the $f$ level empties and the electrons begin to populate the conduction band. Interestingly, the antiferromagnetic order vanishes around $ n_f \sim 0.5$ ($\varepsilon_f \sim 1.5$), \ih{and the system transitions smoothly to a paramagnetic state, where the conduction electrons form a singlet state. The overall behavior and the nature of the phase transition agrees well with the previous density-matrix embedding theory investigation \cite{Yang_2018}.}  If we compare the iPEPS results to the mean-field theory we can see that the mean-field approach reasonably captures the $f$-level occupancy but it significantly overestimates the stability of the antiferromagnetic state, predicting that the paramagnetic solution becomes energetically favorable only around $n_f  \sim 0.1$ ($\varepsilon_f \sim 3.6$). This poorer performance of the mean-field approach can be attributed to the enhanced charge fluctuations, which are more pronounced in the mixed-valence regime than in the localized moment (symmetric case) regime.
\par \ih{In general, we demonstrated that iPEPS provides accurate results for the half-filled case, therefore we can use it to explore other fillings where the quantum Monte Carlo fails.}

\subsection{Away from half filling, $n=1.5$}
\begin{figure}[t]
    \centering
\vspace{0.1cm}
    \begin{overpic}[width=.45\columnwidth]
{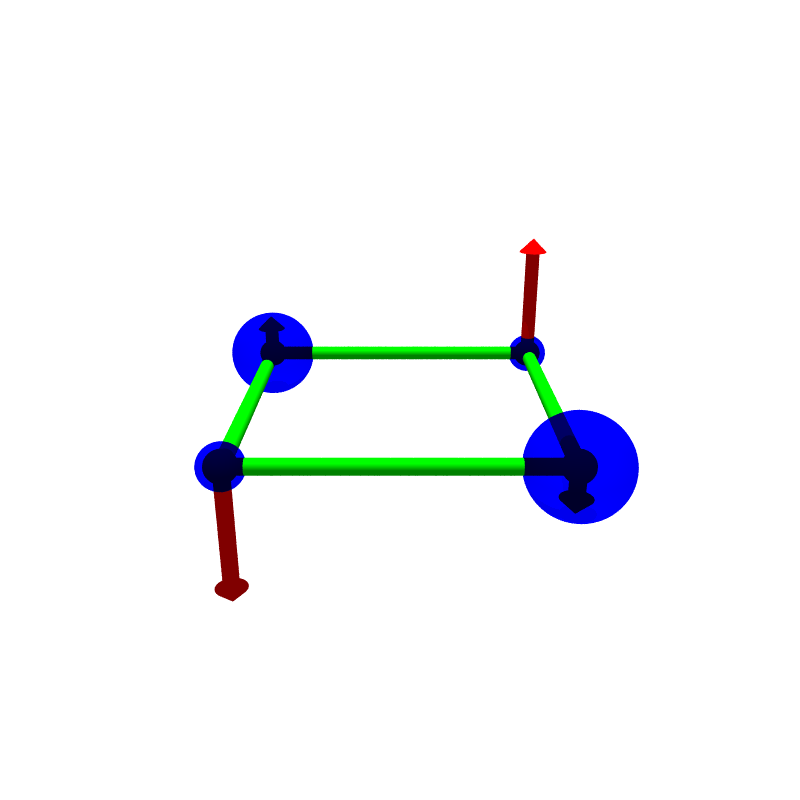}
\put(0,90){(a)}
\put(30,0){\large  diag AF state}
\end{overpic}\hspace{0.5cm}    
\begin{overpic}[width=.45\columnwidth]
{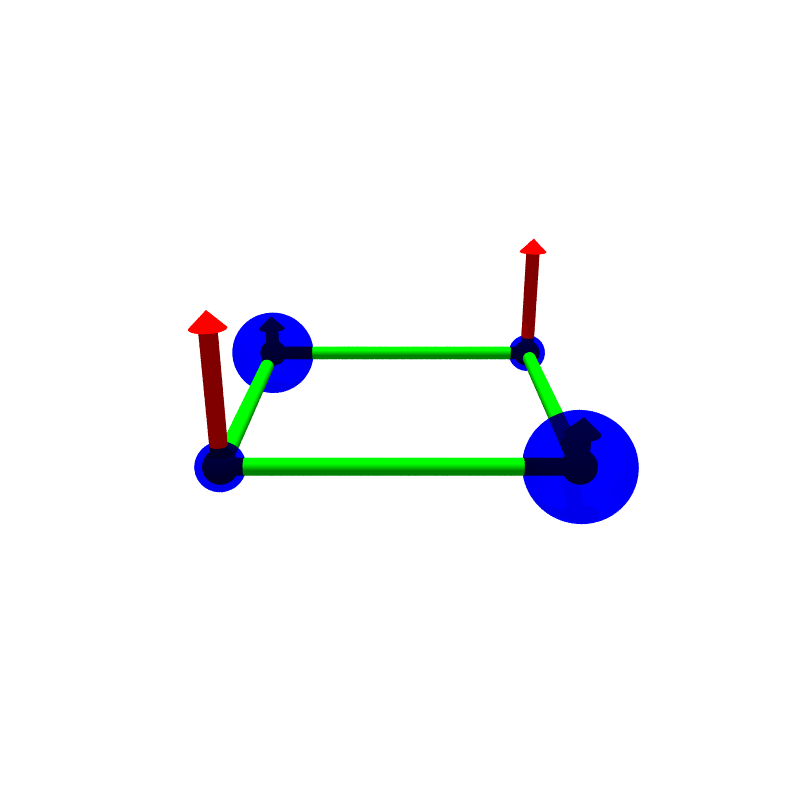}
\put(0,90){(b)}
\put(30,0){\large diag F state}
\end{overpic}  
  \caption{Competing magnetic states that arise in the PAM at $n=1.5$ filling for $\varepsilon_f=-2$, $V=1$ and $U_f=4$. The arrows indicate the magnitude of the onsite magnetizations of the $f$ electrons, which can order antiferromagnetically (a) or ferromagnetically (b). The spheres indicate the strength of onsite correlations, $C$, according to Eq.~(\ref{eq:onsite_corr}). Their radii are chosen as $r\sim \sqrt{C}$ to make the correlations visible at the magnetic sites as well. The magnitudes of the local magnetizations ($m_i^{f}$) on the strongly and weakly magnetized sites are $0.4$ and $0.09$ in the diagonal AF state and $0.41$ and $0.095$ in the diagonal F state at $D=10$.} 
  \label{fig:magnetic_states}
\end{figure}

\begin{figure}[t]
    \centering
\includegraphics[width=\columnwidth]{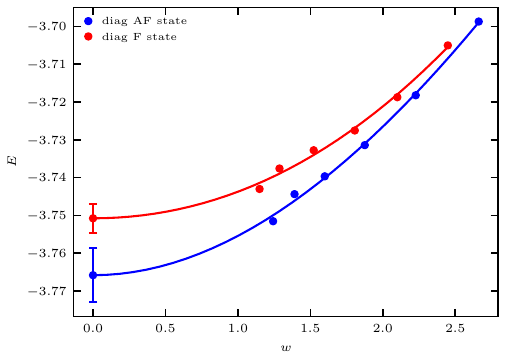}
  \caption{The energy per site for the diag AF and diag F states as a function of the iPEPS cost function, $w$, (divided by the time step of the imaginary time evolution) and as a function of the inverse bond dimension for $U_f=4$, $V=1$, $\varepsilon_f=-2$ and $n=1.5$. The solid lines are third order polynomial fits to the iPEPS data.}
  \label{fig:energy_extrapolation}
\end{figure}

\begin{figure}[!t]
    \centering
\includegraphics[width=\columnwidth]{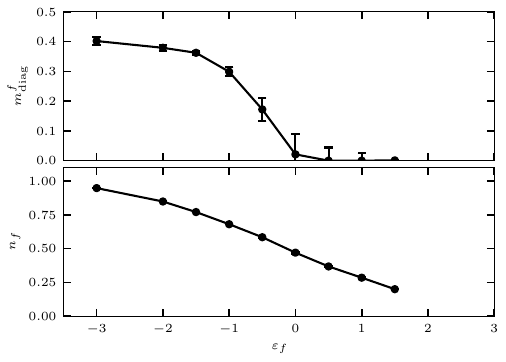}
  \caption{The sublattice magnetization of the $f$ electrons (upper panel) and the $f$ level occupancy (bottom panel) as the function of the $f$-level energy for the parameters $U_f=4$, $V=1$ and $n=1.5$. The symbols are the extrapolated ($D\rightarrow\infty$) iPEPS results. }
  \label{fig:mixed_valence_case_n1.5}
\end{figure}

In the remaining part of this paper, we focus on the case where the system is far from half-filling. This scenario has received less attention due to the limited availability of reliable and unbiased methods. 
Since the ground-state structure is not known a priori, we begin our investigations with a $2 \times 2$ unit cell, allowing four independent tensors at each site \ih{to allow the appearance of more general magnetic structures}. By leveraging the $U(1)$ charge and $U(1)$ spin symmetries, we target specific spin sectors of the unit cell corresponding to $n = 1.5$ filling. Initially, we examine the parameters $U_f = 4$, $\varepsilon_f = -2$, and $V = 1$ \ih{(Mott-Hubbard regime) to ensure that the $f$-level is singly occupied.} Our most striking observation is the emergence of a diagonal stripe phase for the $f$ electrons in the iPEPS simulations. In this phase, antiferromagnetic order occurs along the $[1\bar{1}]$ direction, while the sites along the $[11]$ direction exhibit weak magnetization, as shown in Fig.~\ref{fig:magnetic_states}(a) (AF state). \ih{Note that the symmetry-equivalent states can also be found in iPEPS by using either other random initial states or pinning fields in the beginning of the simulation. In what follows, we use the above convention to characterise the state.}
Since the $f$-level occupancy is $n_f \sim 0.8$ and uniform, the absence of magnetic moments along the $[11]$ direction indicates the presence of strong local correlations. To confirm this, we calculate the local correlator:
\begin{align}
    C=|\langle \hat{n}^f_{j\uparrow} \hat{n}^f_{j\downarrow} \rangle-\langle \hat{n}^f_{j\uparrow} \rangle  \langle \hat{n}^f_{j\downarrow} \rangle|
    \label{eq:onsite_corr}
\end{align}
 and Fig.~\ref{fig:magnetic_states}(a) shows that this correlator is strongly enhanced and very close to its maximal value on the nonmagnetic sites since $\langle \hat{n}^f_{j\uparrow} \hat{n}^f_{j\downarrow} \rangle \sim 0.03$ and $|m_i^f|\sim 0.09$ on the weakly magnetized sites.
 By exploring different spin sectors of the $2 \times 2$ supercell, we found that the state with $S_z^{\rm tot} = 1$ (F state in Fig.~\ref{fig:magnetic_states}(b)) lies slightly above but close in energy to the AF state. To determine which of the competing states is energetically favored in the $D \to \infty$ limit, we extrapolated the energies as a function of the iPEPS cost function. This approach, as suggested in Ref.~\cite{corboz_error}, provides a more accurate estimation compared to extrapolation using the inverse bond dimension. As shown in Fig.~\ref{fig:energy_extrapolation}, a smooth third-order polynomial fit to the data confirms that the AF state, with $E_{AF} = -3.766(7)$, is the ground state, while the F state, with $E_{F} = -3.751(4)$, also belongs to the low-energy manifold.
We also note that for smaller hybridizations, $V \lesssim 0.8$, the energy difference between the competing states becomes smaller, and the error bars begin to overlap. As a result, it is not possible to definitively identify the state with the lowest energy in the error-free limit. Nevertheless, in all cases we examined, the AF state consistently exhibited the lowest variational energy. To characterize the magnetization of the diagonal AF state, $m^{c,f}_{\text{diag }}$, we introduce the quantity
 \begin{align}
     m^{c,f}_{\text{diag}} = \frac{1}{N_{\text{diag}}} \sum_{i \in [1\bar{1}]} (-1)^k m^{c,f}_i,
 \end{align}
 where $k=1$, ($2$) for the diagonal sites of the unit cell with even (odd) indices, for example, $(x,y)=[0,0]$ and $(x,y)=[1,1]$ for $N_{\text{diag}}=2$.
 In Fig.~\ref{fig:mixed_valence_case_n1.5} we show the magnetization of the diagonal AF state as a function of the $f$-level energy.
This  indicates a smooth decrease in magnetization \ih{and a phase transition of second order}, furthermore the appearance of a paramagnetic state for $\varepsilon_f\gtrsim0$, similar to what occurs in the half-filled case. As the Fermi energy in the conduction band is reduced compared to the half-filled case, the $f$-level becomes empty sooner, \cf the bottom panels of Figs.~\ref{fig:mixed_valence_case}-\ref{fig:mixed_valence_case_n1.5}.
A notable feature common to both the $n = 2$ and $n = 1.5$ cases is that the magnetizations vanish around $ n_f  \sim 0.5$ in both scenarios. Additionally, we observe that the conduction electrons remain nearly unpolarized, with $m^{c}_{\text{diag}} \sim \mathcal{O}(10^{-3})$, \ih{in contrast to the half-filled case.} 
 \par Having observed that the system at this filling prefers a diagonal stripe pattern, a natural question arises: does this pattern persist in larger unit cells, or can vertical stripes \ih{or stripes with other periodicity} also emerge?  To address this, we performed exploratory simulations starting from various random initial states using $2 \times 4$ (8 independent tensors) and $4 \times 4$ (16 independent tensors) unit cells. In all cases, the diagonal stripe pattern appeared, supporting our initial findings.
 \par Another important issue is the total spin per unit cell, which can vary continuously in an infinite system. To allow this, we carried out calculations where only $U(1)$ charge conservation was enforced in the iPEPS simulation. This approach, however, limits the maximum bond dimension to $D = 9$ due to the larger dense blocks in the tensors. Starting from various random initial states (including states with $\mathcal{O}(1)$ magnetization per unit cell), we found that the total magnetization decreases as the bond dimension increases, eventually approaching zero. This result confirms our assumption that the total magnetization vanishes in the thermodynamic limit. Furthermore, the lowest-energy states consistently exhibited the diagonal stripe pattern.
 \subsection{Discussion}
  \par Several comments are in order: to discuss the predictions of mean-field theory for this filling and to place our findings in the context of previous studies.
We find that, in the mean-field description, the vertical stripe state -- comprising ferromagnetic lines arranged in an antiferromagnetic pattern -- has the lowest energy when the $f$ level is occupied. However, as outlined above, no signatures of such states were observed in the iPEPS calculations. Consequently, we did not compare the properties of this state with the iPEPS results in Fig.~\ref{fig:mixed_valence_case_n1.5}.
An interesting feature of the mean-field calculation is the presence of a diagonal AF stripe-like solution, although its energy is much higher than that of the vertical stripe state. This discrepancy arises because mean-field theory can lower the energy of the Hubbard term only by inducing magnetic order; it cannot simultaneously suppress double occupancy and maintain paramagnetic behavior at specific sites. As shown in Fig.~\ref{fig:magnetic_states}(a), the strong fluctuations along the $[11]$ diagonal suppress magnetic moments and enhance local correlations. Regarding energetics, we note that although mean-field theory is not a variational approach, even the raw iPEPS energies (without extrapolation) are significantly lower than the mean-field ones for both $n = 2$ and $n = 1.5$.
\par \ih{It is also worth putting these findings in context with previous results in the literature. Mean-field theory on the cubic lattice \cite{spalek_prb}, DMFT on hypercubic lattice \cite{dmft_1997}, variational Monte Carlo as well as \cite{kubo_2015} constrained path quantum Monte Carlo simulations \cite{bonca_1998prb} on the square lattice addressed the ground-state properties of the PAM at $n=1.5$ filling. Mean-field theory, DMFT and variational Monte Carlo predict a ferromagnetic ground state for a wide range of paramaters and occupied $f$-level.} The discrepancy with our findings likely stems from the fact that only conventional antiferromagnetic, paramagnetic, and uniform ferromagnetic states were considered in the variational Monte Carlo and mean-field approaches. The newly identified candidate states from our work cannot be captured by these Ans\"atze, highlighting the need for more general trial wave functions or unit cells in future studies. 
\ih{The results of the constrained path quantum Monte Carlo simulations \cite{bonca_1998prb}, on the other hand, show some similarities with our findings. They found that a resonating spin-density-wave ground state emerges due to the nesting property of the Fermi surface with the wave vectors $\boldsymbol{q}=(\pi,0),(0,\pi)$. This state is characterized by the absence of onsite magnetization, strong antiferromagnetic spin correlations between diagonal sites, and weak correlations between nearest-neighbor sites on a $6 \times 6$ lattice. Since larger system sizes were out of reach for this technique, the question whether this state evolves to long-range order or not in the thermodynamic limit remains inconclusive. The resonating spin-density-wave ground state observed in this finite lattice might be a precursor of the long-range ordered diagonal AF state presented in this paper, although no evidence of its ferromagnetic counterpart was observed in that study.}
\par Finally, we mention that stripe states (beside uniform states) emerge in the two-dimensional Hubbard model around $1/8$ doping as well \cite{leblanc_prx,zheng_2017science} and the vertical one appears to have the lowest energy. There is a strong competition between them but an important difference to our case is that these stripe states host a modulated antiferromagnetic order, while in our case antiferromagnetic and ferromagnetic stripe states compete and we did not find vertical stripes at all in the iPEPS simulations.

\section{Conclusions}
We investigated the ground-state properties of the periodic Anderson model (PAM) on the square lattice at various fillings. Using the iPEPS technique, we obtained unbiased results across a wide range of parameters. First, \ih{to benchmark the applicability of the iPEPS approach}, we examined the symmetric half-filled model and observed that increasing the hybridization leads to the disappearance of antiferromagnetic order. Beyond a critical hybridization value, local Kondo singlets are formed. Our estimate for the critical point is in excellent agreement with quantum Monte Carlo results, underscoring the accuracy of the iPEPS approach.
Next, we studied the system away from the symmetric point by varying the $f$-level energy. We found that antiferromagnetic order persists in the mixed-valence regime but vanishes when the $f$-level occupancy falls below $ n_f \sim 0.5$. A comparison with mean-field theory revealed that, while the mean-field approach provides a qualitatively correct picture, it significantly overestimates the critical point and conduction electron magnetization. This discrepancy becomes even more pronounced in the mixed-valence region.
\par In the second part of the paper, we focused on the doped case with $n = 1.5$ electrons per site. We demonstrated the emergence of novel diagonal (anti-) ferromagnetic stripe states, characterized by strong local correlations along one diagonal. We also highlighted the competition between these states and concluded, based on our energy extrapolations, that the antiferromagnetic stripe state is energetically favored. In contrast, the mean-field calculation incorrectly predicts a vertical stripe state. Overall, our results indicate that the ground state at $n = 1.5$ filling is more complex than those previously proposed in variational Monte Carlo calculations, warranting future studies with more sophisticated trial wave functions.
\par As an outlook, \ih{we mention that it would be interesting to examine how stable the diagonal stripe phase is against doping in the vicinity of $n=1.5$ filling.} Furthermore, we did not explore the possibility of a superconducting ground state in this model, which could provide insights into heavy-fermion superconductors. Another promising extension involves including Hund's coupling via the Kanamori Hamiltonian to investigate how quantum fluctuations modify the phase diagram \cite{koga_hund}. Such studies are challenging for quantum Monte Carlo methods, but as we have shown, iPEPS is capable of providing reliable results for models with large local Hilbert spaces. These problems are therefore within the reach of the iPEPS technique.
\begin{acknowledgments}
The author thanks \"O.~Legeza, L.~Oroszl\'any and P.~Vancs\'o for useful comments and careful reading of the manuscript.
This project was supported by the Hungarian National Research, Development   and   Innovation Office (NKFIH) through Grants No.~FK142985 and No.~K134983, by the Quantum Information National Laboratory of Hungary, as well as by the J\'anos Bolyai Research Scholarship of the Hungarian Academy of Sciences.
We acknowledge the use of the Noctua2 cluster at the Paderborn Center for Parallel Computing (PC$^2$). Some of the data presented here was produced using the \textsc{SyTen} toolkit~\cite{hubig:_syten_toolk,hubig17:_symmet_protec_tensor_networ}. The numerical data that support the findings of this study are available from the author upon reasonable request.
\end{acknowledgments}

\section*{Data availability}
The data that support the findings of this article are openly available \cite{hagymasi_2025_15519541}.

\renewcommand{\theequation}{A\arabic{equation}}

\section*{Appendix: Overview of the iPEPS algorithm}
Here we give a brief overview and further technical details about the iPEPS method employed in the paper. For an in-depth review of the method we refer to Refs.~\cite{cirac_revmod,verstraete2004,nishio2004,cirac_2008prl,corboz_2010prb, nishino_2001,bruognolo_scipost2021}, here we summarize its main aspects. As we outlined in the main text, the iPEPS on a square lattice consists of five-legged tensors. Four of them are auxiliary indices of dimension $D$ that link it to its nearest neighbors, along with a physical index of dimension $d$ representing the local Hilbert space, which is $d=16$ in our model. The bond dimension $D$ governs the accuracy of the variational Ansatz. A ground-state iPEPS simulation unfolds in two main steps. First, the tensor network is optimized to faithfully approximate the ground state of a chosen Hamiltonian. We achieve this optimization via the imaginary-time evolution. Once optimized, the iPEPS enables the calculation of physical properties -- for instance, expectation values of observables. Both stages generally require contracting the two-dimensional tensor network. We discuss these stages in more detail in the following.
\subsection*{Contraction of the tensor network}
\begin{figure}[!t]
    \centering
\includegraphics[width=\columnwidth]{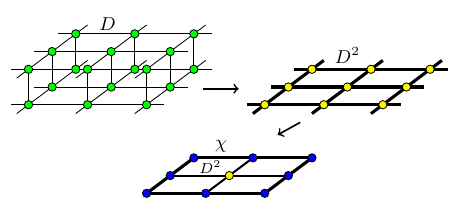}
  \caption{Illustration of the contraction of an iPEPS with bond dimension $D$ using the CTM technique for the calculation of the norm. First, we contract the double-layer iPEPS along the physical indices, then the network of reduced tensors (with bond dimension $D^2$) can be approximated by a set of corner and edge tensors (with bond dimension $\chi$) that encode the information of the infinite environment.}
  \label{fig:ctm}
\end{figure}

    First, we overview the contraction of an iPEPS state, as this is also a prerequisite for the different optimization techniques. The exact contraction of an iPEPS state is exponentially hard, therefore approximate but controlled contraction schemes were developed. We use the corner transfer matrix (CTM) technique \cite{nishino_ctm,orus_ctm}, whose main idea is shown schematically in Fig.~\ref{fig:ctm} for the simplest case, where the iPEPS has a single-site unit cell.
When calculating the norm or the expectation value of a local operator, we need to contract the double-layer network shown in Fig.~\ref{fig:ctm}. The CTM technique makes this contraction feasible by introducing eight fixed-point tensors surrounding the iPEPS tensor in the middle. The accuracy of this technique can be controlled by the bond dimension of the environment tensors, $\chi$. The numerical cost of the CTM technique scales as $\mathcal{O}(\chi^3D^6)$ \cite{bruognolo_scipost2021}, which is $\mathcal{O}(D^{12})$ for our choice with $\chi=D^2$.
\subsection*{Optimization of an iPEPS}
We use the imaginary time evolution for the ground-state search of a given Hamiltonian, $H$. The optimization starts usually from a random initial state $|\phi\rangle$ and for long enough times
\begin{equation}
    |\Psi_{\rm GS}\rangle= \lim_{\beta\rightarrow\infty}\frac{e^{-\beta H}|\phi\rangle}{||e^{-\beta H}|\phi\rangle||}
\end{equation}
the ground state $|\Psi_{\rm GS}\rangle$ is projected out. In order to make this tractable, the time evolution operator is decomposed into two-site gates using the Trotter-Suzuki formula (second order):
\begin{equation}
e^{-\beta H}=e^{-\tau \sum_b{H_b}}=\prod_b e^{-\tau H_b}+\mathcal{O}(\tau^2),  
\end{equation}
where $\tau=\beta/N$ is the time step and the summation extends over the bonds in the supercell. The Trotter error, occurring due to the splitting of the time evolution operator to two-site gates, is well-controlled and can be reduced by making the time steps smaller. After the gate application to an iPEPS bond with bond dimension $D$, it will increase to $d^2D$. To keep the iPEPS state numerically feasible we need to truncate this back to $D$. Unlike matrix-product states, the iPEPS does not have a canonical form because of loops in the lattice, therefore a naive usage of singular value decomposition for the truncation produces suboptimal results. Consequently, a different strategy is utilized to find the truncated tensors. Let us consider a bond $b$ hosting the $A$ and $B$ tensors. After the gate application the new state $|\Psi^{\prime}_{A^{\prime}B^{\prime}}\rangle=e^{-\tau H_b}|\Psi_{AB}\rangle$ needs to be truncated to bond dimension $D$. In the full update this is achieved by finding iteratively the new $\tilde{A}$ and $\tilde{B}$ tensors with bond dimension $D$ that minimize the 
\begin{equation}
    C=|| \ \Psi^{\prime}_{A^{\prime}B^{\prime}}\rangle - |\Psi_{\tilde{A}\tilde{B}}\rangle ||^2
\end{equation}
cost function \cite{corboz_2010prb,orus_ffu_2015prb}. As we mentioned in the main text the quantity $w=C/\tau$ -- which is similar to the truncation error in the density-matrix renormalization group method -- can be used as an error measure to extrapolate the ground-state energy to the error-free limit \cite{corboz_error}. Since the full update requires the calculation of the whole environment in each step, this becomes computationally expensive with increasing the bond dimension. Hence, another approach, the fast-full update \cite{orus_ffu_2015prb} was developed which recycles part of the CTM environment but maintains the accuracy of the update procedure at the same time. We compared both approaches for our Hamiltonian and found that they give identical results up to $D=7$, where the full update was still feasible.  
\bibliography{references}

\begin{thebibliography}{54}%
\makeatletter
\providecommand \@ifxundefined [1]{%
 \@ifx{#1\undefined}
}%
\providecommand \@ifnum [1]{%
 \ifnum #1\expandafter \@firstoftwo
 \else \expandafter \@secondoftwo
 \fi
}%
\providecommand \@ifx [1]{%
 \ifx #1\expandafter \@firstoftwo
 \else \expandafter \@secondoftwo
 \fi
}%
\providecommand \natexlab [1]{#1}%
\providecommand \enquote  [1]{``#1''}%
\providecommand \bibnamefont  [1]{#1}%
\providecommand \bibfnamefont [1]{#1}%
\providecommand \citenamefont [1]{#1}%
\providecommand \href@noop [0]{\@secondoftwo}%
\providecommand \href [0]{\begingroup \@sanitize@url \@href}%
\providecommand \@href[1]{\@@startlink{#1}\@@href}%
\providecommand \@@href[1]{\endgroup#1\@@endlink}%
\providecommand \@sanitize@url [0]{\catcode `\\12\catcode `\$12\catcode `\&12\catcode `\#12\catcode `\^12\catcode `\_12\catcode `\%12\relax}%
\providecommand \@@startlink[1]{}%
\providecommand \@@endlink[0]{}%
\providecommand \url  [0]{\begingroup\@sanitize@url \@url }%
\providecommand \@url [1]{\endgroup\@href {#1}{\urlprefix }}%
\providecommand \urlprefix  [0]{URL }%
\providecommand \Eprint [0]{\href }%
\providecommand \doibase [0]{https://doi.org/}%
\providecommand \selectlanguage [0]{\@gobble}%
\providecommand \bibinfo  [0]{\@secondoftwo}%
\providecommand \bibfield  [0]{\@secondoftwo}%
\providecommand \translation [1]{[#1]}%
\providecommand \BibitemOpen [0]{}%
\providecommand \bibitemStop [0]{}%
\providecommand \bibitemNoStop [0]{.\EOS\space}%
\providecommand \EOS [0]{\spacefactor3000\relax}%
\providecommand \BibitemShut  [1]{\csname bibitem#1\endcsname}%
\let\auto@bib@innerbib\@empty
\bibitem [{\citenamefont {Hewson}(1993)}]{Hewson_1993}%
  \BibitemOpen
  \bibfield  {author} {\bibinfo {author} {\bibfnamefont {A.~C.}\ \bibnamefont {Hewson}},\ }\href@noop {} {\emph {\bibinfo {title} {The Kondo Problem to Heavy Fermions}}},\ Cambridge Studies in Magnetism\ (\bibinfo  {publisher} {Cambridge University Press},\ \bibinfo {year} {1993})\BibitemShut {NoStop}%
\bibitem [{\citenamefont {Fazekas}(1999)}]{fazekas_book_1999}%
  \BibitemOpen
  \bibfield  {author} {\bibinfo {author} {\bibfnamefont {P.}~\bibnamefont {Fazekas}},\ }\href {https://doi.org/10.1142/2945} {\emph {\bibinfo {title} {Lecture Notes on Electron Correlation and Magnetism}}}\ (\bibinfo  {publisher} {WORLD SCIENTIFIC},\ \bibinfo {year} {1999})\ \Eprint {https://arxiv.org/abs/https://www.worldscientific.com/doi/pdf/10.1142/2945} {https://www.worldscientific.com/doi/pdf/10.1142/2945} \BibitemShut {NoStop}%
\bibitem [{\citenamefont {Stewart}(1984)}]{stewart_review}%
  \BibitemOpen
  \bibfield  {author} {\bibinfo {author} {\bibfnamefont {G.~R.}\ \bibnamefont {Stewart}},\ }\bibfield  {title} {\bibinfo {title} {Heavy-fermion systems},\ }\href {https://doi.org/10.1103/RevModPhys.56.755} {\bibfield  {journal} {\bibinfo  {journal} {Rev. Mod. Phys.}\ }\textbf {\bibinfo {volume} {56}},\ \bibinfo {pages} {755} (\bibinfo {year} {1984})}\BibitemShut {NoStop}%
\bibitem [{\citenamefont {Ott}(1994)}]{ott_review}%
  \BibitemOpen
  \bibfield  {author} {\bibinfo {author} {\bibfnamefont {H.}~\bibnamefont {Ott}},\ }\bibfield  {title} {\bibinfo {title} {Magnetism in heavy-electron metals},\ }\href {https://doi.org/10.12693/APhysPolA.85.7} {\bibfield  {journal} {\bibinfo  {journal} {Acta Physica Polonica A}\ }\textbf {\bibinfo {volume} {85}} (\bibinfo {year} {1994})}\BibitemShut {NoStop}%
\bibitem [{\citenamefont {Papavasileiou}\ \emph {et~al.}(2024)\citenamefont {Papavasileiou}, \citenamefont {Menelaou}, \citenamefont {Sarkar}, \citenamefont {Sofer}, \citenamefont {Polavarapu},\ and\ \citenamefont {Mourdikoudis}}]{2d_review}%
  \BibitemOpen
  \bibfield  {author} {\bibinfo {author} {\bibfnamefont {A.~V.}\ \bibnamefont {Papavasileiou}}, \bibinfo {author} {\bibfnamefont {M.}~\bibnamefont {Menelaou}}, \bibinfo {author} {\bibfnamefont {K.~J.}\ \bibnamefont {Sarkar}}, \bibinfo {author} {\bibfnamefont {Z.}~\bibnamefont {Sofer}}, \bibinfo {author} {\bibfnamefont {L.}~\bibnamefont {Polavarapu}},\ and\ \bibinfo {author} {\bibfnamefont {S.}~\bibnamefont {Mourdikoudis}},\ }\bibfield  {title} {\bibinfo {title} {Ferromagnetic elements in two-dimensional materials: 2d magnets and beyond},\ }\href {https://doi.org/https://doi.org/10.1002/adfm.202309046} {\bibfield  {journal} {\bibinfo  {journal} {Advanced Functional Materials}\ }\textbf {\bibinfo {volume} {34}},\ \bibinfo {pages} {2309046} (\bibinfo {year} {2024})}\BibitemShut {NoStop}%
\bibitem [{\citenamefont {Noce}(2006)}]{noce_review}%
  \BibitemOpen
  \bibfield  {author} {\bibinfo {author} {\bibfnamefont {C.}~\bibnamefont {Noce}},\ }\bibfield  {title} {\bibinfo {title} {The periodic anderson model: Symmetry-based results and some exact solutions},\ }\href {https://doi.org/https://doi.org/10.1016/j.physrep.2006.05.003} {\bibfield  {journal} {\bibinfo  {journal} {Physics Reports}\ }\textbf {\bibinfo {volume} {431}},\ \bibinfo {pages} {173} (\bibinfo {year} {2006})}\BibitemShut {NoStop}%
\bibitem [{\citenamefont {Gul\'acsi}\ and\ \citenamefont {Vollhardt}(2003)}]{gulacsi_prl}%
  \BibitemOpen
  \bibfield  {author} {\bibinfo {author} {\bibfnamefont {Z.}~\bibnamefont {Gul\'acsi}}\ and\ \bibinfo {author} {\bibfnamefont {D.}~\bibnamefont {Vollhardt}},\ }\bibfield  {title} {\bibinfo {title} {Exact insulating and conducting ground states of a periodic anderson model in three dimensions},\ }\href {https://doi.org/10.1103/PhysRevLett.91.186401} {\bibfield  {journal} {\bibinfo  {journal} {Phys. Rev. Lett.}\ }\textbf {\bibinfo {volume} {91}},\ \bibinfo {pages} {186401} (\bibinfo {year} {2003})}\BibitemShut {NoStop}%
\bibitem [{\citenamefont {Gurin}\ and\ \citenamefont {Gul\'acsi}(2001)}]{gurin_2001prb}%
  \BibitemOpen
  \bibfield  {author} {\bibinfo {author} {\bibfnamefont {P.}~\bibnamefont {Gurin}}\ and\ \bibinfo {author} {\bibfnamefont {Z.}~\bibnamefont {Gul\'acsi}},\ }\bibfield  {title} {\bibinfo {title} {Exact solutions for the periodic anderson model in two dimensions: A non-fermi-liquid state in the normal phase},\ }\href {https://doi.org/10.1103/PhysRevB.64.045118} {\bibfield  {journal} {\bibinfo  {journal} {Phys. Rev. B}\ }\textbf {\bibinfo {volume} {64}},\ \bibinfo {pages} {045118} (\bibinfo {year} {2001})}\BibitemShut {NoStop}%
\bibitem [{\citenamefont {Guerrero}\ and\ \citenamefont {Noack}(2001)}]{noack_prb2001}%
  \BibitemOpen
  \bibfield  {author} {\bibinfo {author} {\bibfnamefont {M.}~\bibnamefont {Guerrero}}\ and\ \bibinfo {author} {\bibfnamefont {R.~M.}\ \bibnamefont {Noack}},\ }\bibfield  {title} {\bibinfo {title} {Ferromagnetism and phase separation in one-dimensional $d\ensuremath{-}p$ and periodic anderson models},\ }\href {https://doi.org/10.1103/PhysRevB.63.144423} {\bibfield  {journal} {\bibinfo  {journal} {Phys. Rev. B}\ }\textbf {\bibinfo {volume} {63}},\ \bibinfo {pages} {144423} (\bibinfo {year} {2001})}\BibitemShut {NoStop}%
\bibitem [{\citenamefont {Bertussi}\ \emph {et~al.}(2011)\citenamefont {Bertussi}, \citenamefont {Neto}, \citenamefont {Rappoport}, \citenamefont {Malvezzi},\ and\ \citenamefont {dos Santos}}]{santos_2011prb}%
  \BibitemOpen
  \bibfield  {author} {\bibinfo {author} {\bibfnamefont {P.~R.}\ \bibnamefont {Bertussi}}, \bibinfo {author} {\bibfnamefont {M.~B.~S.}\ \bibnamefont {Neto}}, \bibinfo {author} {\bibfnamefont {T.~G.}\ \bibnamefont {Rappoport}}, \bibinfo {author} {\bibfnamefont {A.~L.}\ \bibnamefont {Malvezzi}},\ and\ \bibinfo {author} {\bibfnamefont {R.~R.}\ \bibnamefont {dos Santos}},\ }\bibfield  {title} {\bibinfo {title} {Incommensurate spin-density-wave and metal-insulator transition in the one-dimensional periodic anderson model},\ }\href {https://doi.org/10.1103/PhysRevB.84.075156} {\bibfield  {journal} {\bibinfo  {journal} {Phys. Rev. B}\ }\textbf {\bibinfo {volume} {84}},\ \bibinfo {pages} {075156} (\bibinfo {year} {2011})}\BibitemShut {NoStop}%
\bibitem [{\citenamefont {Hagym\'asi}\ \emph {et~al.}(2014)\citenamefont {Hagym\'asi}, \citenamefont {S\'olyom},\ and\ \citenamefont {Legeza}}]{hagymasi_2014prb}%
  \BibitemOpen
  \bibfield  {author} {\bibinfo {author} {\bibfnamefont {I.}~\bibnamefont {Hagym\'asi}}, \bibinfo {author} {\bibfnamefont {J.}~\bibnamefont {S\'olyom}},\ and\ \bibinfo {author} {\bibfnamefont {O.}~\bibnamefont {Legeza}},\ }\bibfield  {title} {\bibinfo {title} {Interorbital interaction in the one-dimensional periodic anderson model: Density matrix renormalization group study},\ }\href {https://doi.org/10.1103/PhysRevB.90.125137} {\bibfield  {journal} {\bibinfo  {journal} {Phys. Rev. B}\ }\textbf {\bibinfo {volume} {90}},\ \bibinfo {pages} {125137} (\bibinfo {year} {2014})}\BibitemShut {NoStop}%
\bibitem [{\citenamefont {Veki\ifmmode~\acute{c}\else \'{c}\fi{}}\ \emph {et~al.}(1995)\citenamefont {Veki\ifmmode~\acute{c}\else \'{c}\fi{}}, \citenamefont {Cannon}, \citenamefont {Scalapino}, \citenamefont {Scalettar},\ and\ \citenamefont {Sugar}}]{vekic_prl}%
  \BibitemOpen
  \bibfield  {author} {\bibinfo {author} {\bibfnamefont {M.}~\bibnamefont {Veki\ifmmode~\acute{c}\else \'{c}\fi{}}}, \bibinfo {author} {\bibfnamefont {J.~W.}\ \bibnamefont {Cannon}}, \bibinfo {author} {\bibfnamefont {D.~J.}\ \bibnamefont {Scalapino}}, \bibinfo {author} {\bibfnamefont {R.~T.}\ \bibnamefont {Scalettar}},\ and\ \bibinfo {author} {\bibfnamefont {R.~L.}\ \bibnamefont {Sugar}},\ }\bibfield  {title} {\bibinfo {title} {Competition between antiferromagnetic order and spin-liquid behavior in the two-dimensional periodic anderson model at half filling},\ }\href {https://doi.org/10.1103/PhysRevLett.74.2367} {\bibfield  {journal} {\bibinfo  {journal} {Phys. Rev. Lett.}\ }\textbf {\bibinfo {volume} {74}},\ \bibinfo {pages} {2367} (\bibinfo {year} {1995})}\BibitemShut {NoStop}%
\bibitem [{\citenamefont {Watanabe}\ and\ \citenamefont {Ogata}(2009)}]{watanabe_2009}%
  \BibitemOpen
  \bibfield  {author} {\bibinfo {author} {\bibfnamefont {H.}~\bibnamefont {Watanabe}}\ and\ \bibinfo {author} {\bibfnamefont {M.}~\bibnamefont {Ogata}},\ }\bibfield  {title} {\bibinfo {title} {Fermi-surface reconstruction in the periodic anderson model},\ }\href@noop {} {\bibfield  {journal} {\bibinfo  {journal} {Journal of the Physical Society of Japan}\ }\textbf {\bibinfo {volume} {78}},\ \bibinfo {pages} {024715} (\bibinfo {year} {2009})}\BibitemShut {NoStop}%
\bibitem [{\citenamefont {Kubo}(2015)}]{kubo_2015}%
  \BibitemOpen
  \bibfield  {author} {\bibinfo {author} {\bibfnamefont {K.}~\bibnamefont {Kubo}},\ }\bibfield  {title} {\bibinfo {title} {Lifshitz transitions in magnetic phases of the periodic anderson model},\ }\href@noop {} {\bibfield  {journal} {\bibinfo  {journal} {Journal of the Physical Society of Japan}\ }\textbf {\bibinfo {volume} {84}},\ \bibinfo {pages} {094702} (\bibinfo {year} {2015})}\BibitemShut {NoStop}%
\bibitem [{\citenamefont {Sch\"afer}\ \emph {et~al.}(2019)\citenamefont {Sch\"afer}, \citenamefont {Katanin}, \citenamefont {Kitatani}, \citenamefont {Toschi},\ and\ \citenamefont {Held}}]{schafer_2019}%
  \BibitemOpen
  \bibfield  {author} {\bibinfo {author} {\bibfnamefont {T.}~\bibnamefont {Sch\"afer}}, \bibinfo {author} {\bibfnamefont {A.~A.}\ \bibnamefont {Katanin}}, \bibinfo {author} {\bibfnamefont {M.}~\bibnamefont {Kitatani}}, \bibinfo {author} {\bibfnamefont {A.}~\bibnamefont {Toschi}},\ and\ \bibinfo {author} {\bibfnamefont {K.}~\bibnamefont {Held}},\ }\bibfield  {title} {\bibinfo {title} {Quantum criticality in the two-dimensional periodic anderson model},\ }\href {https://doi.org/10.1103/PhysRevLett.122.227201} {\bibfield  {journal} {\bibinfo  {journal} {Phys. Rev. Lett.}\ }\textbf {\bibinfo {volume} {122}},\ \bibinfo {pages} {227201} (\bibinfo {year} {2019})}\BibitemShut {NoStop}%
\bibitem [{\citenamefont {Doradzi\ifmmode~\acute{n}\else \'{n}\fi{}ski}\ and\ \citenamefont {Spa\l{}ek}(1998)}]{spalek_prb}%
  \BibitemOpen
  \bibfield  {author} {\bibinfo {author} {\bibfnamefont {R.}~\bibnamefont {Doradzi\ifmmode~\acute{n}\else \'{n}\fi{}ski}}\ and\ \bibinfo {author} {\bibfnamefont {J.}~\bibnamefont {Spa\l{}ek}},\ }\bibfield  {title} {\bibinfo {title} {Mean-field magnetic phase diagram of the periodic anderson model with the kondo-compensated phases},\ }\href {https://doi.org/10.1103/PhysRevB.58.3293} {\bibfield  {journal} {\bibinfo  {journal} {Phys. Rev. B}\ }\textbf {\bibinfo {volume} {58}},\ \bibinfo {pages} {3293} (\bibinfo {year} {1998})}\BibitemShut {NoStop}%
\bibitem [{\citenamefont {Meyer}\ and\ \citenamefont {Nolting}(2000)}]{nolting_prb}%
  \BibitemOpen
  \bibfield  {author} {\bibinfo {author} {\bibfnamefont {D.}~\bibnamefont {Meyer}}\ and\ \bibinfo {author} {\bibfnamefont {W.}~\bibnamefont {Nolting}},\ }\bibfield  {title} {\bibinfo {title} {Dynamical mean-field study of ferromagnetism in the periodic anderson model},\ }\href {https://doi.org/10.1103/PhysRevB.62.5657} {\bibfield  {journal} {\bibinfo  {journal} {Phys. Rev. B}\ }\textbf {\bibinfo {volume} {62}},\ \bibinfo {pages} {5657} (\bibinfo {year} {2000})}\BibitemShut {NoStop}%
\bibitem [{\citenamefont {Koga}\ \emph {et~al.}(2008{\natexlab{a}})\citenamefont {Koga}, \citenamefont {Kawakami}, \citenamefont {Peters},\ and\ \citenamefont {Pruschke}}]{koga_prb}%
  \BibitemOpen
  \bibfield  {author} {\bibinfo {author} {\bibfnamefont {A.}~\bibnamefont {Koga}}, \bibinfo {author} {\bibfnamefont {N.}~\bibnamefont {Kawakami}}, \bibinfo {author} {\bibfnamefont {R.}~\bibnamefont {Peters}},\ and\ \bibinfo {author} {\bibfnamefont {T.}~\bibnamefont {Pruschke}},\ }\bibfield  {title} {\bibinfo {title} {Quantum phase transitions in the extended periodic anderson model},\ }\href {https://doi.org/10.1103/PhysRevB.77.045120} {\bibfield  {journal} {\bibinfo  {journal} {Phys. Rev. B}\ }\textbf {\bibinfo {volume} {77}},\ \bibinfo {pages} {045120} (\bibinfo {year} {2008}{\natexlab{a}})}\BibitemShut {NoStop}%
\bibitem [{\citenamefont {De~Leo}\ \emph {et~al.}(2008)\citenamefont {De~Leo}, \citenamefont {Civelli},\ and\ \citenamefont {Kotliar}}]{kotliar_2008prl}%
  \BibitemOpen
  \bibfield  {author} {\bibinfo {author} {\bibfnamefont {L.}~\bibnamefont {De~Leo}}, \bibinfo {author} {\bibfnamefont {M.}~\bibnamefont {Civelli}},\ and\ \bibinfo {author} {\bibfnamefont {G.}~\bibnamefont {Kotliar}},\ }\bibfield  {title} {\bibinfo {title} {$t=0$ heavy-fermion quantum critical point as an orbital-selective mott transition},\ }\href {https://doi.org/10.1103/PhysRevLett.101.256404} {\bibfield  {journal} {\bibinfo  {journal} {Phys. Rev. Lett.}\ }\textbf {\bibinfo {volume} {101}},\ \bibinfo {pages} {256404} (\bibinfo {year} {2008})}\BibitemShut {NoStop}%
\bibitem [{\citenamefont {Gleis}\ \emph {et~al.}(2024)\citenamefont {Gleis}, \citenamefont {Lee}, \citenamefont {Kotliar},\ and\ \citenamefont {von Delft}}]{vondelft_2024prx}%
  \BibitemOpen
  \bibfield  {author} {\bibinfo {author} {\bibfnamefont {A.}~\bibnamefont {Gleis}}, \bibinfo {author} {\bibfnamefont {S.-S.~B.}\ \bibnamefont {Lee}}, \bibinfo {author} {\bibfnamefont {G.}~\bibnamefont {Kotliar}},\ and\ \bibinfo {author} {\bibfnamefont {J.}~\bibnamefont {von Delft}},\ }\bibfield  {title} {\bibinfo {title} {Emergent properties of the periodic anderson model: A high-resolution, real-frequency study of heavy-fermion quantum criticality},\ }\href {https://doi.org/10.1103/PhysRevX.14.041036} {\bibfield  {journal} {\bibinfo  {journal} {Phys. Rev. X}\ }\textbf {\bibinfo {volume} {14}},\ \bibinfo {pages} {041036} (\bibinfo {year} {2024})}\BibitemShut {NoStop}%
\bibitem [{\citenamefont {Troyer}\ and\ \citenamefont {Wiese}(2005)}]{troyer_prl}%
  \BibitemOpen
  \bibfield  {author} {\bibinfo {author} {\bibfnamefont {M.}~\bibnamefont {Troyer}}\ and\ \bibinfo {author} {\bibfnamefont {U.-J.}\ \bibnamefont {Wiese}},\ }\bibfield  {title} {\bibinfo {title} {Computational complexity and fundamental limitations to fermionic quantum monte carlo simulations},\ }\href {https://doi.org/10.1103/PhysRevLett.94.170201} {\bibfield  {journal} {\bibinfo  {journal} {Phys. Rev. Lett.}\ }\textbf {\bibinfo {volume} {94}},\ \bibinfo {pages} {170201} (\bibinfo {year} {2005})}\BibitemShut {NoStop}%
\bibitem [{\citenamefont {Bon\ifmmode~\check{c}\else \v{c}\fi{}a}\ and\ \citenamefont {Gubernatis}(1998)}]{bonca_1998prb}%
  \BibitemOpen
  \bibfield  {author} {\bibinfo {author} {\bibfnamefont {J.}~\bibnamefont {Bon\ifmmode~\check{c}\else \v{c}\fi{}a}}\ and\ \bibinfo {author} {\bibfnamefont {J.~E.}\ \bibnamefont {Gubernatis}},\ }\bibfield  {title} {\bibinfo {title} {Effects of doping on spin correlations in the periodic anderson model},\ }\href {https://doi.org/10.1103/PhysRevB.58.6992} {\bibfield  {journal} {\bibinfo  {journal} {Phys. Rev. B}\ }\textbf {\bibinfo {volume} {58}},\ \bibinfo {pages} {6992} (\bibinfo {year} {1998})}\BibitemShut {NoStop}%
\bibitem [{\citenamefont {Batista}\ \emph {et~al.}(2001)\citenamefont {Batista}, \citenamefont {Bon\ifmmode~\check{c}\else \v{c}\fi{}a},\ and\ \citenamefont {Gubernatis}}]{batista_2001prb}%
  \BibitemOpen
  \bibfield  {author} {\bibinfo {author} {\bibfnamefont {C.~D.}\ \bibnamefont {Batista}}, \bibinfo {author} {\bibfnamefont {J.}~\bibnamefont {Bon\ifmmode~\check{c}\else \v{c}\fi{}a}},\ and\ \bibinfo {author} {\bibfnamefont {J.~E.}\ \bibnamefont {Gubernatis}},\ }\bibfield  {title} {\bibinfo {title} {Ferromagnetism in the two-dimensional periodic anderson model},\ }\href {https://doi.org/10.1103/PhysRevB.63.184428} {\bibfield  {journal} {\bibinfo  {journal} {Phys. Rev. B}\ }\textbf {\bibinfo {volume} {63}},\ \bibinfo {pages} {184428} (\bibinfo {year} {2001})}\BibitemShut {NoStop}%
\bibitem [{\citenamefont {Fazekas}\ and\ \citenamefont {Brandow}(1987)}]{fazekas_1987}%
  \BibitemOpen
  \bibfield  {author} {\bibinfo {author} {\bibfnamefont {P.}~\bibnamefont {Fazekas}}\ and\ \bibinfo {author} {\bibfnamefont {B.~H.}\ \bibnamefont {Brandow}},\ }\bibfield  {title} {\bibinfo {title} {Application of the gutzwiller method to the periodic anderson model},\ }\href {https://doi.org/10.1088/0031-8949/36/5/008} {\bibfield  {journal} {\bibinfo  {journal} {Physica Scripta}\ }\textbf {\bibinfo {volume} {36}},\ \bibinfo {pages} {809} (\bibinfo {year} {1987})}\BibitemShut {NoStop}%
\bibitem [{\citenamefont {Gebhard}(1991)}]{gebhard_prb1991}%
  \BibitemOpen
  \bibfield  {author} {\bibinfo {author} {\bibfnamefont {F.}~\bibnamefont {Gebhard}},\ }\bibfield  {title} {\bibinfo {title} {Equivalence of variational and slave-boson mean-field treatments of the periodic anderson model},\ }\href {https://doi.org/10.1103/PhysRevB.44.992} {\bibfield  {journal} {\bibinfo  {journal} {Phys. Rev. B}\ }\textbf {\bibinfo {volume} {44}},\ \bibinfo {pages} {992} (\bibinfo {year} {1991})}\BibitemShut {NoStop}%
\bibitem [{\citenamefont {Itai}\ and\ \citenamefont {Fazekas}(1996)}]{itai_prb}%
  \BibitemOpen
  \bibfield  {author} {\bibinfo {author} {\bibfnamefont {K.}~\bibnamefont {Itai}}\ and\ \bibinfo {author} {\bibfnamefont {P.}~\bibnamefont {Fazekas}},\ }\bibfield  {title} {\bibinfo {title} {Interaction effect in the kondo energy of the periodic anderson-hubbard model},\ }\href {https://doi.org/10.1103/PhysRevB.54.R752} {\bibfield  {journal} {\bibinfo  {journal} {Phys. Rev. B}\ }\textbf {\bibinfo {volume} {54}},\ \bibinfo {pages} {R752} (\bibinfo {year} {1996})}\BibitemShut {NoStop}%
\bibitem [{\citenamefont {Hagym\'asi}\ \emph {et~al.}(2012)\citenamefont {Hagym\'asi}, \citenamefont {Itai},\ and\ \citenamefont {S\'olyom}}]{hagymasi_2012prb}%
  \BibitemOpen
  \bibfield  {author} {\bibinfo {author} {\bibfnamefont {I.}~\bibnamefont {Hagym\'asi}}, \bibinfo {author} {\bibfnamefont {K.}~\bibnamefont {Itai}},\ and\ \bibinfo {author} {\bibfnamefont {J.}~\bibnamefont {S\'olyom}},\ }\bibfield  {title} {\bibinfo {title} {Periodic anderson model with correlated conduction electrons: Variational and exact diagonalization study},\ }\href {https://doi.org/10.1103/PhysRevB.85.235116} {\bibfield  {journal} {\bibinfo  {journal} {Phys. Rev. B}\ }\textbf {\bibinfo {volume} {85}},\ \bibinfo {pages} {235116} (\bibinfo {year} {2012})}\BibitemShut {NoStop}%
\bibitem [{\citenamefont {Yang}\ and\ \citenamefont {Chen}(2018)}]{Yang_2018}%
  \BibitemOpen
  \bibfield  {author} {\bibinfo {author} {\bibfnamefont {J.-W.}\ \bibnamefont {Yang}}\ and\ \bibinfo {author} {\bibfnamefont {Q.-N.}\ \bibnamefont {Chen}},\ }\bibfield  {title} {\bibinfo {title} {Phase diagram, correlations, and quantum critical point in the periodic anderson model*},\ }\href@noop {} {\bibfield  {journal} {\bibinfo  {journal} {Chinese Physics B}\ }\textbf {\bibinfo {volume} {27}},\ \bibinfo {pages} {037101} (\bibinfo {year} {2018})}\BibitemShut {NoStop}%
\bibitem [{\citenamefont {Cirac}\ \emph {et~al.}(2021)\citenamefont {Cirac}, \citenamefont {P\'erez-Garc\'{\i}a}, \citenamefont {Schuch},\ and\ \citenamefont {Verstraete}}]{cirac_revmod}%
  \BibitemOpen
  \bibfield  {author} {\bibinfo {author} {\bibfnamefont {J.~I.}\ \bibnamefont {Cirac}}, \bibinfo {author} {\bibfnamefont {D.}~\bibnamefont {P\'erez-Garc\'{\i}a}}, \bibinfo {author} {\bibfnamefont {N.}~\bibnamefont {Schuch}},\ and\ \bibinfo {author} {\bibfnamefont {F.}~\bibnamefont {Verstraete}},\ }\bibfield  {title} {\bibinfo {title} {Matrix product states and projected entangled pair states: Concepts, symmetries, theorems},\ }\href {https://doi.org/10.1103/RevModPhys.93.045003} {\bibfield  {journal} {\bibinfo  {journal} {Rev. Mod. Phys.}\ }\textbf {\bibinfo {volume} {93}},\ \bibinfo {pages} {045003} (\bibinfo {year} {2021})}\BibitemShut {NoStop}%
\bibitem [{\citenamefont {Verstraete}\ and\ \citenamefont {Cirac}(2004)}]{verstraete2004}%
  \BibitemOpen
  \bibfield  {author} {\bibinfo {author} {\bibfnamefont {F.}~\bibnamefont {Verstraete}}\ and\ \bibinfo {author} {\bibfnamefont {J.~I.}\ \bibnamefont {Cirac}},\ }\href {https://arxiv.org/abs/cond-mat/0407066} {\bibinfo {title} {Renormalization algorithms for quantum-many body systems in two and higher dimensions}} (\bibinfo {year} {2004}),\ \Eprint {https://arxiv.org/abs/cond-mat/0407066} {arXiv:cond-mat/0407066 [cond-mat.str-el]} \BibitemShut {NoStop}%
\bibitem [{\citenamefont {Nishio}\ \emph {et~al.}(2004)\citenamefont {Nishio}, \citenamefont {Maeshima}, \citenamefont {Gendiar},\ and\ \citenamefont {Nishino}}]{nishio2004}%
  \BibitemOpen
  \bibfield  {author} {\bibinfo {author} {\bibfnamefont {Y.}~\bibnamefont {Nishio}}, \bibinfo {author} {\bibfnamefont {N.}~\bibnamefont {Maeshima}}, \bibinfo {author} {\bibfnamefont {A.}~\bibnamefont {Gendiar}},\ and\ \bibinfo {author} {\bibfnamefont {T.}~\bibnamefont {Nishino}},\ }\href {https://arxiv.org/abs/cond-mat/0401115} {\bibinfo {title} {Tensor product variational formulation for quantum systems}} (\bibinfo {year} {2004}),\ \Eprint {https://arxiv.org/abs/cond-mat/0401115} {arXiv:cond-mat/0401115 [cond-mat.stat-mech]} \BibitemShut {NoStop}%
\bibitem [{\citenamefont {Jordan}\ \emph {et~al.}(2008)\citenamefont {Jordan}, \citenamefont {Or\'us}, \citenamefont {Vidal}, \citenamefont {Verstraete},\ and\ \citenamefont {Cirac}}]{cirac_2008prl}%
  \BibitemOpen
  \bibfield  {author} {\bibinfo {author} {\bibfnamefont {J.}~\bibnamefont {Jordan}}, \bibinfo {author} {\bibfnamefont {R.}~\bibnamefont {Or\'us}}, \bibinfo {author} {\bibfnamefont {G.}~\bibnamefont {Vidal}}, \bibinfo {author} {\bibfnamefont {F.}~\bibnamefont {Verstraete}},\ and\ \bibinfo {author} {\bibfnamefont {J.~I.}\ \bibnamefont {Cirac}},\ }\bibfield  {title} {\bibinfo {title} {Classical simulation of infinite-size quantum lattice systems in two spatial dimensions},\ }\href {https://doi.org/10.1103/PhysRevLett.101.250602} {\bibfield  {journal} {\bibinfo  {journal} {Phys. Rev. Lett.}\ }\textbf {\bibinfo {volume} {101}},\ \bibinfo {pages} {250602} (\bibinfo {year} {2008})}\BibitemShut {NoStop}%
\bibitem [{\citenamefont {Corboz}\ \emph {et~al.}(2010)\citenamefont {Corboz}, \citenamefont {Or\'us}, \citenamefont {Bauer},\ and\ \citenamefont {Vidal}}]{corboz_2010prb}%
  \BibitemOpen
  \bibfield  {author} {\bibinfo {author} {\bibfnamefont {P.}~\bibnamefont {Corboz}}, \bibinfo {author} {\bibfnamefont {R.}~\bibnamefont {Or\'us}}, \bibinfo {author} {\bibfnamefont {B.}~\bibnamefont {Bauer}},\ and\ \bibinfo {author} {\bibfnamefont {G.}~\bibnamefont {Vidal}},\ }\bibfield  {title} {\bibinfo {title} {Simulation of strongly correlated fermions in two spatial dimensions with fermionic projected entangled-pair states},\ }\href {https://doi.org/10.1103/PhysRevB.81.165104} {\bibfield  {journal} {\bibinfo  {journal} {Phys. Rev. B}\ }\textbf {\bibinfo {volume} {81}},\ \bibinfo {pages} {165104} (\bibinfo {year} {2010})}\BibitemShut {NoStop}%
\bibitem [{\citenamefont {Nishino}\ \emph {et~al.}(2001)\citenamefont {Nishino}, \citenamefont {Hieida}, \citenamefont {Okunishi}, \citenamefont {Maeshima}, \citenamefont {Akutsu},\ and\ \citenamefont {Gendiar}}]{nishino_2001}%
  \BibitemOpen
  \bibfield  {author} {\bibinfo {author} {\bibfnamefont {T.}~\bibnamefont {Nishino}}, \bibinfo {author} {\bibfnamefont {Y.}~\bibnamefont {Hieida}}, \bibinfo {author} {\bibfnamefont {K.}~\bibnamefont {Okunishi}}, \bibinfo {author} {\bibfnamefont {N.}~\bibnamefont {Maeshima}}, \bibinfo {author} {\bibfnamefont {Y.}~\bibnamefont {Akutsu}},\ and\ \bibinfo {author} {\bibfnamefont {A.}~\bibnamefont {Gendiar}},\ }\bibfield  {title} {\bibinfo {title} {Two-dimensional tensor product variational formulation},\ }\href {https://doi.org/10.1143/PTP.105.409} {\bibfield  {journal} {\bibinfo  {journal} {Progress of Theoretical Physics}\ }\textbf {\bibinfo {volume} {105}},\ \bibinfo {pages} {409} (\bibinfo {year} {2001})},\ \Eprint {https://arxiv.org/abs/https://academic.oup.com/ptp/article-pdf/105/3/409/5191613/105-3-409.pdf} {https://academic.oup.com/ptp/article-pdf/105/3/409/5191613/105-3-409.pdf} \BibitemShut {NoStop}%
\bibitem [{\citenamefont {Peschke}\ \emph {et~al.}(2022)\citenamefont {Peschke}, \citenamefont {Ponsioen},\ and\ \citenamefont {Corboz}}]{peschke_2022prb}%
  \BibitemOpen
  \bibfield  {author} {\bibinfo {author} {\bibfnamefont {M.}~\bibnamefont {Peschke}}, \bibinfo {author} {\bibfnamefont {B.}~\bibnamefont {Ponsioen}},\ and\ \bibinfo {author} {\bibfnamefont {P.}~\bibnamefont {Corboz}},\ }\bibfield  {title} {\bibinfo {title} {Competing states in the two-dimensional frustrated kondo-necklace model},\ }\href {https://doi.org/10.1103/PhysRevB.106.205140} {\bibfield  {journal} {\bibinfo  {journal} {Phys. Rev. B}\ }\textbf {\bibinfo {volume} {106}},\ \bibinfo {pages} {205140} (\bibinfo {year} {2022})}\BibitemShut {NoStop}%
\bibitem [{\citenamefont {Corboz}\ and\ \citenamefont {Mila}(2014)}]{corboz_2014prl}%
  \BibitemOpen
  \bibfield  {author} {\bibinfo {author} {\bibfnamefont {P.}~\bibnamefont {Corboz}}\ and\ \bibinfo {author} {\bibfnamefont {F.}~\bibnamefont {Mila}},\ }\bibfield  {title} {\bibinfo {title} {Crystals of bound states in the magnetization plateaus of the shastry-sutherland model},\ }\href {https://doi.org/10.1103/PhysRevLett.112.147203} {\bibfield  {journal} {\bibinfo  {journal} {Phys. Rev. Lett.}\ }\textbf {\bibinfo {volume} {112}},\ \bibinfo {pages} {147203} (\bibinfo {year} {2014})}\BibitemShut {NoStop}%
\bibitem [{\citenamefont {Liao}\ \emph {et~al.}(2017)\citenamefont {Liao}, \citenamefont {Xie}, \citenamefont {Chen}, \citenamefont {Liu}, \citenamefont {Xie}, \citenamefont {Huang}, \citenamefont {Normand},\ and\ \citenamefont {Xiang}}]{liao_2017prl}%
  \BibitemOpen
  \bibfield  {author} {\bibinfo {author} {\bibfnamefont {H.~J.}\ \bibnamefont {Liao}}, \bibinfo {author} {\bibfnamefont {Z.~Y.}\ \bibnamefont {Xie}}, \bibinfo {author} {\bibfnamefont {J.}~\bibnamefont {Chen}}, \bibinfo {author} {\bibfnamefont {Z.~Y.}\ \bibnamefont {Liu}}, \bibinfo {author} {\bibfnamefont {H.~D.}\ \bibnamefont {Xie}}, \bibinfo {author} {\bibfnamefont {R.~Z.}\ \bibnamefont {Huang}}, \bibinfo {author} {\bibfnamefont {B.}~\bibnamefont {Normand}},\ and\ \bibinfo {author} {\bibfnamefont {T.}~\bibnamefont {Xiang}},\ }\bibfield  {title} {\bibinfo {title} {Gapless spin-liquid ground state in the $s=1/2$ kagome antiferromagnet},\ }\href {https://doi.org/10.1103/PhysRevLett.118.137202} {\bibfield  {journal} {\bibinfo  {journal} {Phys. Rev. Lett.}\ }\textbf {\bibinfo {volume} {118}},\ \bibinfo {pages} {137202} (\bibinfo {year} {2017})}\BibitemShut {NoStop}%
\bibitem [{\citenamefont {Schmoll}\ \emph {et~al.}(2023)\citenamefont {Schmoll}, \citenamefont {Kshetrimayum}, \citenamefont {Naumann}, \citenamefont {Eisert},\ and\ \citenamefont {Iqbal}}]{schmoll_2023prb}%
  \BibitemOpen
  \bibfield  {author} {\bibinfo {author} {\bibfnamefont {P.}~\bibnamefont {Schmoll}}, \bibinfo {author} {\bibfnamefont {A.}~\bibnamefont {Kshetrimayum}}, \bibinfo {author} {\bibfnamefont {J.}~\bibnamefont {Naumann}}, \bibinfo {author} {\bibfnamefont {J.}~\bibnamefont {Eisert}},\ and\ \bibinfo {author} {\bibfnamefont {Y.}~\bibnamefont {Iqbal}},\ }\bibfield  {title} {\bibinfo {title} {Tensor network study of the spin-$\frac{1}{2}$ heisenberg antiferromagnet on the shuriken lattice},\ }\href {https://doi.org/10.1103/PhysRevB.107.064406} {\bibfield  {journal} {\bibinfo  {journal} {Phys. Rev. B}\ }\textbf {\bibinfo {volume} {107}},\ \bibinfo {pages} {064406} (\bibinfo {year} {2023})}\BibitemShut {NoStop}%
\bibitem [{\citenamefont {Zheng}\ \emph {et~al.}(2017)\citenamefont {Zheng}, \citenamefont {Chung}, \citenamefont {Corboz}, \citenamefont {Ehlers}, \citenamefont {Qin}, \citenamefont {Noack}, \citenamefont {Shi}, \citenamefont {White}, \citenamefont {Zhang},\ and\ \citenamefont {Chan}}]{zheng_2017science}%
  \BibitemOpen
  \bibfield  {author} {\bibinfo {author} {\bibfnamefont {B.-X.}\ \bibnamefont {Zheng}}, \bibinfo {author} {\bibfnamefont {C.-M.}\ \bibnamefont {Chung}}, \bibinfo {author} {\bibfnamefont {P.}~\bibnamefont {Corboz}}, \bibinfo {author} {\bibfnamefont {G.}~\bibnamefont {Ehlers}}, \bibinfo {author} {\bibfnamefont {M.-P.}\ \bibnamefont {Qin}}, \bibinfo {author} {\bibfnamefont {R.~M.}\ \bibnamefont {Noack}}, \bibinfo {author} {\bibfnamefont {H.}~\bibnamefont {Shi}}, \bibinfo {author} {\bibfnamefont {S.~R.}\ \bibnamefont {White}}, \bibinfo {author} {\bibfnamefont {S.}~\bibnamefont {Zhang}},\ and\ \bibinfo {author} {\bibfnamefont {G.~K.-L.}\ \bibnamefont {Chan}},\ }\bibfield  {title} {\bibinfo {title} {Stripe order in the underdoped region of the two-dimensional hubbard model},\ }\href {https://doi.org/10.1126/science.aam7127} {\bibfield  {journal} {\bibinfo  {journal} {Science}\ }\textbf {\bibinfo {volume} {358}},\ \bibinfo {pages} {1155} (\bibinfo {year} {2017})},\ \Eprint
  {https://arxiv.org/abs/https://www.science.org/doi/pdf/10.1126/science.aam7127} {https://www.science.org/doi/pdf/10.1126/science.aam7127} \BibitemShut {NoStop}%
\bibitem [{\citenamefont {Corboz}\ \emph {et~al.}(2014)\citenamefont {Corboz}, \citenamefont {Rice},\ and\ \citenamefont {Troyer}}]{corboz_tJ_2014prl}%
  \BibitemOpen
  \bibfield  {author} {\bibinfo {author} {\bibfnamefont {P.}~\bibnamefont {Corboz}}, \bibinfo {author} {\bibfnamefont {T.~M.}\ \bibnamefont {Rice}},\ and\ \bibinfo {author} {\bibfnamefont {M.}~\bibnamefont {Troyer}},\ }\bibfield  {title} {\bibinfo {title} {Competing states in the $t$-$j$ model: Uniform $d$-wave state versus stripe state},\ }\href {https://doi.org/10.1103/PhysRevLett.113.046402} {\bibfield  {journal} {\bibinfo  {journal} {Phys. Rev. Lett.}\ }\textbf {\bibinfo {volume} {113}},\ \bibinfo {pages} {046402} (\bibinfo {year} {2014})}\BibitemShut {NoStop}%
\bibitem [{\citenamefont {Lee}\ \emph {et~al.}(2020)\citenamefont {Lee}, \citenamefont {Kaneko}, \citenamefont {Chern}, \citenamefont {Okubo}, \citenamefont {Yamaji}, \citenamefont {Kawashima},\ and\ \citenamefont {Kim}}]{Lee2020}%
  \BibitemOpen
  \bibfield  {author} {\bibinfo {author} {\bibfnamefont {H.-Y.}\ \bibnamefont {Lee}}, \bibinfo {author} {\bibfnamefont {R.}~\bibnamefont {Kaneko}}, \bibinfo {author} {\bibfnamefont {L.~E.}\ \bibnamefont {Chern}}, \bibinfo {author} {\bibfnamefont {T.}~\bibnamefont {Okubo}}, \bibinfo {author} {\bibfnamefont {Y.}~\bibnamefont {Yamaji}}, \bibinfo {author} {\bibfnamefont {N.}~\bibnamefont {Kawashima}},\ and\ \bibinfo {author} {\bibfnamefont {Y.~B.}\ \bibnamefont {Kim}},\ }\bibfield  {title} {\bibinfo {title} {Magnetic field induced quantum phases in a tensor network study of kitaev magnets},\ }\href {https://doi.org/10.1038/s41467-020-15320-x} {\bibfield  {journal} {\bibinfo  {journal} {Nature Communications}\ }\textbf {\bibinfo {volume} {11}},\ \bibinfo {pages} {1639} (\bibinfo {year} {2020})}\BibitemShut {NoStop}%
\bibitem [{\citenamefont {Bruognolo}\ \emph {et~al.}(2021)\citenamefont {Bruognolo}, \citenamefont {Li}, \citenamefont {von Delft},\ and\ \citenamefont {Weichselbaum}}]{bruognolo_scipost2021}%
  \BibitemOpen
  \bibfield  {author} {\bibinfo {author} {\bibfnamefont {B.}~\bibnamefont {Bruognolo}}, \bibinfo {author} {\bibfnamefont {J.-W.}\ \bibnamefont {Li}}, \bibinfo {author} {\bibfnamefont {J.}~\bibnamefont {von Delft}},\ and\ \bibinfo {author} {\bibfnamefont {A.}~\bibnamefont {Weichselbaum}},\ }\bibfield  {title} {\bibinfo {title} {{A beginner's guide to non-abelian iPEPS for correlated fermions}},\ }\href {https://doi.org/10.21468/SciPostPhysLectNotes.25} {\bibfield  {journal} {\bibinfo  {journal} {SciPost Phys. Lect. Notes}\ ,\ \bibinfo {pages} {25}} (\bibinfo {year} {2021})}\BibitemShut {NoStop}%
\bibitem [{\citenamefont {Phien}\ \emph {et~al.}(2015)\citenamefont {Phien}, \citenamefont {Bengua}, \citenamefont {Tuan}, \citenamefont {Corboz},\ and\ \citenamefont {Or\'us}}]{orus_ffu_2015prb}%
  \BibitemOpen
  \bibfield  {author} {\bibinfo {author} {\bibfnamefont {H.~N.}\ \bibnamefont {Phien}}, \bibinfo {author} {\bibfnamefont {J.~A.}\ \bibnamefont {Bengua}}, \bibinfo {author} {\bibfnamefont {H.~D.}\ \bibnamefont {Tuan}}, \bibinfo {author} {\bibfnamefont {P.}~\bibnamefont {Corboz}},\ and\ \bibinfo {author} {\bibfnamefont {R.}~\bibnamefont {Or\'us}},\ }\bibfield  {title} {\bibinfo {title} {Infinite projected entangled pair states algorithm improved: Fast full update and gauge fixing},\ }\href {https://doi.org/10.1103/PhysRevB.92.035142} {\bibfield  {journal} {\bibinfo  {journal} {Phys. Rev. B}\ }\textbf {\bibinfo {volume} {92}},\ \bibinfo {pages} {035142} (\bibinfo {year} {2015})}\BibitemShut {NoStop}%
\bibitem [{\citenamefont {Nishino}\ and\ \citenamefont {Okunishi}(1996)}]{nishino_ctm}%
  \BibitemOpen
  \bibfield  {author} {\bibinfo {author} {\bibfnamefont {T.}~\bibnamefont {Nishino}}\ and\ \bibinfo {author} {\bibfnamefont {K.}~\bibnamefont {Okunishi}},\ }\bibfield  {title} {\bibinfo {title} {Corner transfer matrix renormalization group method},\ }\href {https://doi.org/10.1143/JPSJ.65.891} {\bibfield  {journal} {\bibinfo  {journal} {Journal of the Physical Society of Japan}\ }\textbf {\bibinfo {volume} {65}},\ \bibinfo {pages} {891} (\bibinfo {year} {1996})},\ \Eprint {https://arxiv.org/abs/https://doi.org/10.1143/JPSJ.65.891} {https://doi.org/10.1143/JPSJ.65.891} \BibitemShut {NoStop}%
\bibitem [{\citenamefont {Or\'us}\ and\ \citenamefont {Vidal}(2009)}]{orus_ctm}%
  \BibitemOpen
  \bibfield  {author} {\bibinfo {author} {\bibfnamefont {R.}~\bibnamefont {Or\'us}}\ and\ \bibinfo {author} {\bibfnamefont {G.}~\bibnamefont {Vidal}},\ }\bibfield  {title} {\bibinfo {title} {Simulation of two-dimensional quantum systems on an infinite lattice revisited: Corner transfer matrix for tensor contraction},\ }\href {https://doi.org/10.1103/PhysRevB.80.094403} {\bibfield  {journal} {\bibinfo  {journal} {Phys. Rev. B}\ }\textbf {\bibinfo {volume} {80}},\ \bibinfo {pages} {094403} (\bibinfo {year} {2009})}\BibitemShut {NoStop}%
\bibitem [{\citenamefont {Corboz}(2016)}]{corboz_error}%
  \BibitemOpen
  \bibfield  {author} {\bibinfo {author} {\bibfnamefont {P.}~\bibnamefont {Corboz}},\ }\bibfield  {title} {\bibinfo {title} {Improved energy extrapolation with infinite projected entangled-pair states applied to the two-dimensional hubbard model},\ }\href {https://doi.org/10.1103/PhysRevB.93.045116} {\bibfield  {journal} {\bibinfo  {journal} {Phys. Rev. B}\ }\textbf {\bibinfo {volume} {93}},\ \bibinfo {pages} {045116} (\bibinfo {year} {2016})}\BibitemShut {NoStop}%
\bibitem [{\citenamefont {Hagym\'asi}\ \emph {et~al.}(2013)\citenamefont {Hagym\'asi}, \citenamefont {Itai},\ and\ \citenamefont {S\'olyom}}]{hagymasi_2013prb}%
  \BibitemOpen
  \bibfield  {author} {\bibinfo {author} {\bibfnamefont {I.}~\bibnamefont {Hagym\'asi}}, \bibinfo {author} {\bibfnamefont {K.}~\bibnamefont {Itai}},\ and\ \bibinfo {author} {\bibfnamefont {J.}~\bibnamefont {S\'olyom}},\ }\bibfield  {title} {\bibinfo {title} {Quantum criticality and first-order transitions in the extended periodic anderson model},\ }\href {https://doi.org/10.1103/PhysRevB.87.125146} {\bibfield  {journal} {\bibinfo  {journal} {Phys. Rev. B}\ }\textbf {\bibinfo {volume} {87}},\ \bibinfo {pages} {125146} (\bibinfo {year} {2013})}\BibitemShut {NoStop}%
\bibitem [{\citenamefont {Assaad}(1999)}]{assaad_klm_prl}%
  \BibitemOpen
  \bibfield  {author} {\bibinfo {author} {\bibfnamefont {F.~F.}\ \bibnamefont {Assaad}},\ }\bibfield  {title} {\bibinfo {title} {Quantum monte carlo simulations of the half-filled two-dimensional kondo lattice model},\ }\href {https://doi.org/10.1103/PhysRevLett.83.796} {\bibfield  {journal} {\bibinfo  {journal} {Phys. Rev. Lett.}\ }\textbf {\bibinfo {volume} {83}},\ \bibinfo {pages} {796} (\bibinfo {year} {1999})}\BibitemShut {NoStop}%
\bibitem [{\citenamefont {Tahvildar-Zadeh}\ \emph {et~al.}(1997)\citenamefont {Tahvildar-Zadeh}, \citenamefont {Jarrell},\ and\ \citenamefont {Freericks}}]{dmft_1997}%
  \BibitemOpen
  \bibfield  {author} {\bibinfo {author} {\bibfnamefont {A.~N.}\ \bibnamefont {Tahvildar-Zadeh}}, \bibinfo {author} {\bibfnamefont {M.}~\bibnamefont {Jarrell}},\ and\ \bibinfo {author} {\bibfnamefont {J.~K.}\ \bibnamefont {Freericks}},\ }\bibfield  {title} {\bibinfo {title} {Protracted screening in the periodic anderson model},\ }\href {https://doi.org/10.1103/PhysRevB.55.R3332} {\bibfield  {journal} {\bibinfo  {journal} {Phys. Rev. B}\ }\textbf {\bibinfo {volume} {55}},\ \bibinfo {pages} {R3332} (\bibinfo {year} {1997})}\BibitemShut {NoStop}%
\bibitem [{\citenamefont {LeBlanc}\ \emph {et~al.}(2015)\citenamefont {LeBlanc}, \citenamefont {Antipov}, \citenamefont {Becca}, \citenamefont {Bulik}, \citenamefont {Chan}, \citenamefont {Chung}, \citenamefont {Deng}, \citenamefont {Ferrero}, \citenamefont {Henderson}, \citenamefont {Jim\'enez-Hoyos}, \citenamefont {Kozik}, \citenamefont {Liu}, \citenamefont {Millis}, \citenamefont {Prokof'ev}, \citenamefont {Qin}, \citenamefont {Scuseria}, \citenamefont {Shi}, \citenamefont {Svistunov}, \citenamefont {Tocchio}, \citenamefont {Tupitsyn}, \citenamefont {White}, \citenamefont {Zhang}, \citenamefont {Zheng}, \citenamefont {Zhu},\ and\ \citenamefont {Gull}}]{leblanc_prx}%
  \BibitemOpen
  \bibfield  {author} {\bibinfo {author} {\bibfnamefont {J.~P.~F.}\ \bibnamefont {LeBlanc}}, \bibinfo {author} {\bibfnamefont {A.~E.}\ \bibnamefont {Antipov}}, \bibinfo {author} {\bibfnamefont {F.}~\bibnamefont {Becca}}, \bibinfo {author} {\bibfnamefont {I.~W.}\ \bibnamefont {Bulik}}, \bibinfo {author} {\bibfnamefont {G.~K.-L.}\ \bibnamefont {Chan}}, \bibinfo {author} {\bibfnamefont {C.-M.}\ \bibnamefont {Chung}}, \bibinfo {author} {\bibfnamefont {Y.}~\bibnamefont {Deng}}, \bibinfo {author} {\bibfnamefont {M.}~\bibnamefont {Ferrero}}, \bibinfo {author} {\bibfnamefont {T.~M.}\ \bibnamefont {Henderson}}, \bibinfo {author} {\bibfnamefont {C.~A.}\ \bibnamefont {Jim\'enez-Hoyos}}, \bibinfo {author} {\bibfnamefont {E.}~\bibnamefont {Kozik}}, \bibinfo {author} {\bibfnamefont {X.-W.}\ \bibnamefont {Liu}}, \bibinfo {author} {\bibfnamefont {A.~J.}\ \bibnamefont {Millis}}, \bibinfo {author} {\bibfnamefont {N.~V.}\ \bibnamefont {Prokof'ev}}, \bibinfo {author} {\bibfnamefont {M.}~\bibnamefont {Qin}}, \bibinfo {author}
  {\bibfnamefont {G.~E.}\ \bibnamefont {Scuseria}}, \bibinfo {author} {\bibfnamefont {H.}~\bibnamefont {Shi}}, \bibinfo {author} {\bibfnamefont {B.~V.}\ \bibnamefont {Svistunov}}, \bibinfo {author} {\bibfnamefont {L.~F.}\ \bibnamefont {Tocchio}}, \bibinfo {author} {\bibfnamefont {I.~S.}\ \bibnamefont {Tupitsyn}}, \bibinfo {author} {\bibfnamefont {S.~R.}\ \bibnamefont {White}}, \bibinfo {author} {\bibfnamefont {S.}~\bibnamefont {Zhang}}, \bibinfo {author} {\bibfnamefont {B.-X.}\ \bibnamefont {Zheng}}, \bibinfo {author} {\bibfnamefont {Z.}~\bibnamefont {Zhu}},\ and\ \bibinfo {author} {\bibfnamefont {E.}~\bibnamefont {Gull}} (\bibinfo {collaboration} {Simons Collaboration on the Many-Electron Problem}),\ }\bibfield  {title} {\bibinfo {title} {Solutions of the two-dimensional hubbard model: Benchmarks and results from a wide range of numerical algorithms},\ }\href {https://doi.org/10.1103/PhysRevX.5.041041} {\bibfield  {journal} {\bibinfo  {journal} {Phys. Rev. X}\ }\textbf {\bibinfo {volume} {5}},\ \bibinfo
  {pages} {041041} (\bibinfo {year} {2015})}\BibitemShut {NoStop}%
\bibitem [{\citenamefont {Koga}\ \emph {et~al.}(2008{\natexlab{b}})\citenamefont {Koga}, \citenamefont {Kawakami}, \citenamefont {Peters},\ and\ \citenamefont {Pruschke}}]{koga_hund}%
  \BibitemOpen
  \bibfield  {author} {\bibinfo {author} {\bibfnamefont {A.}~\bibnamefont {Koga}}, \bibinfo {author} {\bibfnamefont {N.}~\bibnamefont {Kawakami}}, \bibinfo {author} {\bibfnamefont {R.}~\bibnamefont {Peters}},\ and\ \bibinfo {author} {\bibfnamefont {T.}~\bibnamefont {Pruschke}},\ }\bibfield  {title} {\bibinfo {title} {Magnetic properties of the extended periodic anderson model},\ }\href {https://doi.org/10.1143/JPSJ.77.033704} {\bibfield  {journal} {\bibinfo  {journal} {Journal of the Physical Society of Japan}\ }\textbf {\bibinfo {volume} {77}},\ \bibinfo {pages} {033704} (\bibinfo {year} {2008}{\natexlab{b}})},\ \Eprint {https://arxiv.org/abs/https://doi.org/10.1143/JPSJ.77.033704} {https://doi.org/10.1143/JPSJ.77.033704} \BibitemShut {NoStop}%
\bibitem [{\citenamefont {Hubig}\ \emph {et~al.}()\citenamefont {Hubig}, \citenamefont {Lachenmaier}, \citenamefont {Linden}, \citenamefont {Reinhard}, \citenamefont {Stenzel}, \citenamefont {Swoboda},\ and\ \citenamefont {Grundner}}]{hubig:_syten_toolk}%
  \BibitemOpen
  \bibfield  {author} {\bibinfo {author} {\bibfnamefont {C.}~\bibnamefont {Hubig}}, \bibinfo {author} {\bibfnamefont {F.}~\bibnamefont {Lachenmaier}}, \bibinfo {author} {\bibfnamefont {N.-O.}\ \bibnamefont {Linden}}, \bibinfo {author} {\bibfnamefont {T.}~\bibnamefont {Reinhard}}, \bibinfo {author} {\bibfnamefont {L.}~\bibnamefont {Stenzel}}, \bibinfo {author} {\bibfnamefont {A.}~\bibnamefont {Swoboda}},\ and\ \bibinfo {author} {\bibfnamefont {M.}~\bibnamefont {Grundner}},\ }\href {https://syten.eu} {\bibinfo {title} {The \textsc{SyTen} toolkit}}\BibitemShut {NoStop}%
\bibitem [{\citenamefont {Hubig}(2017)}]{hubig17:_symmet_protec_tensor_networ}%
  \BibitemOpen
  \bibfield  {author} {\bibinfo {author} {\bibfnamefont {C.}~\bibnamefont {Hubig}},\ }\emph {\bibinfo {title} {Symmetry-Protected Tensor Networks}},\ \href {https://edoc.ub.uni-muenchen.de/21348/} {Ph.D. thesis},\ \bibinfo  {school} {LMU M\"unchen} (\bibinfo {year} {2017})\BibitemShut {NoStop}%
\bibitem [{\citenamefont {Hagymasi}(2025)}]{hagymasi_2025_15519541}%
  \BibitemOpen
  \bibfield  {author} {\bibinfo {author} {\bibfnamefont {I.}~\bibnamefont {Hagymasi}},\ }\bibfield  {title} {\bibinfo {title} {Raw data for the manuscript "magnetic phases of the periodic anderson model in two dimensions"},\ }\href {https://doi.org/10.5281/zenodo.15519541} {10.5281/zenodo.15519541} (\bibinfo {year} {2025})\BibitemShut {NoStop}%
\end{thebibliography}%
\end{document}